\documentclass[aps,
raggedbottom,
prb,
showpacs,
nobalancelastpage,
groupedaddress]{revtex4}
\usepackage{graphicx}
\usepackage{amsmath}
\usepackage{amsfonts}
\usepackage{amssymb}
\usepackage{pifont}
\usepackage{hyperref}
\usepackage{units}
\usepackage{eufrak}

\newcommand{\da}{\dagger}
\newcommand{\ve}[1]{\ensuremath{\mathbf{#1}}}

\newcommand{\cre}[2]{\ensuremath{#1_{#2}^\da}}
\newcommand{\ann}[2]{\ensuremath{#1_{#2}^{ }}}
\newcommand{\abso}[1]{\lvert #1 \rvert}
\newcommand{\p}{\prime}
\newcommand{\com}[1]{[#1]}

\newcommand{\sgn}{\mathrm{sgn}}
\newcommand{\rt}{\tilde{r}}
\newcommand{\Rt}{\tilde{R}}
\newcommand{\Lt}{\tilde{L}}

\newcommand{\tc}{\tilde{c}}
\newcommand{\tU}{\tilde{U}}

\newcommand{\mlead}{\mathrm{leads}}
\newcommand{\mE}{\mathcal{E}}
\newcommand{\mF}{\mathcal{F}}
\newcommand{\mN}{\mathcal{N}}
\newcommand{\mL}{\mathcal{L}}
\newcommand{\mP}{\mathcal{P}}
\newcommand{\ket}[1]{\rvert #1 \rangle}
\newcommand{\tro}[2]{\ensuremath{\mathrm{Tr}_{#1}\{#2\}}}
\newcommand{\tr}[1]{\ensuremath{\mathrm{Tr}\{#1\}}}
\newcommand{\veg}[1]{\ensuremath{\boldsymbol{#1}}}
\newcommand{\cdwnt}{\ensuremath{\text{c-DWCNT}}}
\newcommand{\idwnt}{\ensuremath{\text{i-DWCNT}}}
\newcommand{\mmax}{\ensuremath{\mathrm{max}}}
\newcommand{\mmin}{\ensuremath{\mathrm{min}}}

\allowdisplaybreaks[1]

\begin{document}

\title{Transport properties of double-walled carbon nanotube quantum
  dots}

\author{Shidong Wang and Milena Grifoni} \affiliation{Theoretische
  Physik, Universit\"at Regensburg, 93040 Regensburg, Germany.}

\date{\today}

\begin{abstract}
The transport properties of quantum dot (QD) systems based on
double-walled carbon nanotube (DWCNT) are investigated. The interplay
between microscopic structure and strong Coulomb interaction is
treated within a bosonization framework. The linear and nonlinear
$G$-$V$-$V_g$ characteristics of the QD system is  calculated by
starting from  the Liouville equation for the reduced density
matrix. Depending on the intershell couplings, an 8-electron
periodicity of the Coulomb blockade peak spacing in the case of
commensurate DWCNT QDs and a 4-electron periodicity in the
incommensurate case are predicted. The contribution of excited states
of DWCNTs to the nonlinear transport is investigated as well.
\end{abstract}


\maketitle

\section{Introduction}
\label{sec:intro}

After being discovered in 1991~\cite{ijima:nature1991}, carbon
nanotubes (CNTs) have been widely used in 
nano-devices because of their unique
properties~\cite{saito:1998, charlier:677, loiseau-nt:2006}. CNTs may be either
single-walled (SWCNT) or multi-walled (MWCNT) depending on the number
of graphene sheets wrapped into concentric cylinders. Due to the
quasi-one-dimensional characters of their electronic structures, long SWCNTs
exhibit Luttinger-liquid
behavior~\cite{egger:prl1997, egger:epjb1998, kane:prl1997,
  bockrath:nature1999, postma2001cns}. 
SWCNT quantum dot (QD) systems
have also been fabricated, which consist of finite length SWCNTs
weakly connected to the source and drain leads and capacitively
coupled to
a gate electrode~\cite{liang:prl2002, moriyama:prl2005,
  sapmaz:prb2005, sapmaz2006qdc}. At low bias, the SWCNT QD systems
show Coulomb blockade behavior because of the strong Coulomb
interactions in the QDs and the poor transparencies of
the contacts between the QD and the leads~\cite{grabert1991set}.
Because of the short lengths of SWCNTs,
the addition energy 
needed to add an extra electron to the QD depends on both the Coulomb
interaction and on the energy level spacing. Unlike the traditional
two-dimensional semiconductor 
QD systems with irregular Coulomb blockade patterns, which have to be
understood statistically~\cite{alhassid:rmp2000}, QDs based on
SWCNTs 
show regular Coulomb
blockade patterns, which originate from the
regular electronic structure of the SWCNTs.
Because of the spin degeneracy of 
two bands crossing at the Fermi points in metallic SWCNTs, 
the stability diagrams of SWCNT QD systems exhibit a
4-electron periodicity of the Coulomb diamond
sizes~\cite{liang:prl2002, cobden:prl2002, moriyama:prl2005, sapmaz:prb2005,
  sapmaz2006qdc}. 
The stability diagrams of the SWCNT QD systems
have been
explained by using the mean-field theory developed in
Ref.~\onlinecite{oreg:prl2000} which includes a nonzero 
exchange energy~\cite{liang:prl2002,  moriyama:prl2005, sapmaz:prb2005,
  sapmaz2006qdc}. Recently, the energy spectrum  of SWCNT QD has been
calculated in Refs.~\onlinecite{mayrhofer:prb2006}
and~\onlinecite{mayrhofer:epjb2007} beyond mean field. For QD systems with
moderate-to-large 
radius SWCNTs, the exchange energy can be 
ignored~\cite{mayrhofer}, and the stability 
diagrams can also be quantitatively explained within a bosonization
approach~\cite{mayrhofer:prb2006, mayrhofer:epjb2007}.  By suitable
choice of parameters these
theories can reproduce the \emph{same} low bias
spectra of SWCNT QDs, and only the excitations measured at high
bias are predicted differently by a mean-field approach or by a
bosonization method because of the
different treatment of the Coulomb
interaction~\cite{mayrhofer:epjb2007}. Although 
the excitations of SWCNT QDs have already been
measured~\cite{sapmaz:prb2005}, the quality and the range of the
measured excitations cannot be used to determine the validity of
these two methods and further experiments are needed.

So far, the properties of MWCNT QD systems have not been fully
explored~\cite{buitellar:prl2002}. 
The experiment in Ref.~\onlinecite{buitellar:prl2002}
showed that the stability diagrams of MWCNT QD systems  
have a 4-electron periodicity of the Coulomb diamond
sizes.
The simplest 
MWCNT QD is the one formed by a double-walled carbon nanotube (DWCNT),
which consists of two concentric shells. 
Depending on the ratio between the unit cell lengths of the two shells,
a DWCNT may be either 
commensurate (c-DWCNT), if the ratio is a 
commensurate number, or incommensurate (i-DWCNT), if the ratio
is an incommensurate number. It has been shown that the effective
intershell coupling depends on the type of DWCNTs. At low energies,
that is, near the Fermi energy, the effective intershell coupling is
negligible in i-DWCNTs but large in
c-DWCNTs while it cannot be ignored in both type of DWCNTs at high
energies~\cite{yoon:prb2002, 
  roche:prb2001, triozon:prb2004, wang:2005}. 
Both types of DWCNTs with long lengths
can be described by Luttinger liquid theory when Coulomb interactions
are included~\cite{egger:prl1999, 
  wang:2005}. 
Because of their intermediate-to-large radii, we expect
that the exchange energy may be ignored in DWCNTs. Therefore,
the bosonization approach, which includes forward scattering processes
exactly, can be used to describe the
properties of DWCNT QD systems as well.

In this paper, we consider a QD system formed by a finite length DWCNT
with
two metallic shells, where we include all forward scattering
processes. The bosonization approach enables exact diagonalization of
the interacting DWCNT Hamiltonian. 
Finally, the linear and nonlinear transport
properties of the system are investigated by
solving the Liouville equation for the reduced density matrix
to lowest order in the coupling to the leads. 

The paper is organized as follows. In Sec.~\ref{sec:model-method},
the Hamiltonian of a DWCNT QD system is derived. The energy spectrum of a
finite length DWCNT with strong Coulomb interactions and 
open boundary conditions is then obtained. Transport properties of DWCNT
QDs are calculated in Sec.~\ref{sec:dynamics-qd}. The results for the
linear and 
nonlinear conductances of both c-DWCNT and i-DWCNT QDs are presented in
Sec.~\ref{sec:results}. Finally, the conclusion is drawn in
Sec.~\ref{sec:conclusions}.

\section{Model and method}
\label{sec:model-method}

\begin{figure}
 \includegraphics[width=0.4\textwidth]{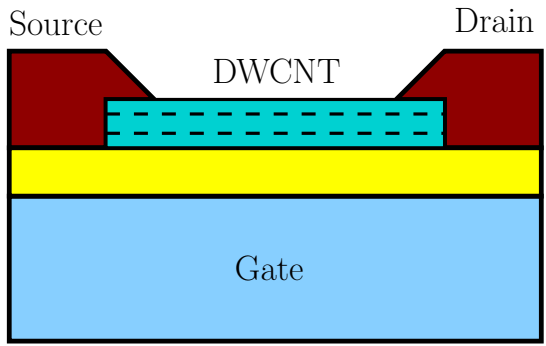} 
  \caption{(Color online) Schematic experimental setup of a
    double-walled carbon 
    nanotube (DWCNT) quantum dot (QD) system. A finite length DWCNT is
    deposited on a substrate and weakly connected to the source and drain
    leads through its outer shell. A gate electrode is capacitively
    coupled to the DWCNT QD 
    and controls the electrochemical potential in the QD.  The dashed
    lines denote the inner shell in the DWCNT.
    }
  \label{fig:qd-exp}
\end{figure}

As schematically shown in Fig.~\ref{fig:qd-exp}, the QD system
consists of a  
DWCNT with two metallic shells deposited on a
substrate. 
The source and drain leads are connected to the outer shell of the
DWCNT. The segment of the DWCNT
(of about several hundred nanometers long)
between two leads forms a QD. A gate electrode is capacitively
coupled to the QD and controls the electrochemical potential in it. As
we are only interested in the Coulomb blockade regime,  
we assume that the QD is weakly contacted 
to two leads, that is, the transparencies of the contacts are very poor
and the conductance of the QD system is much smaller than the
conductance quantum $2e^2/h$. 
The Hamiltonian of the whole system can be separated into several parts,
\begin{equation}
  \label{eq:H-total1}
  H = H_{\mlead} + H_{QD} + H_T + H_g,
\end{equation}
where $H_{QD}$ is the Hamiltonian of the QD system and its explicit
form will be derived in the following subsection. The source ($s$) and
drain ($d$) leads are described by Fermi gases of non-interacting
quasi-particles and the Hamiltonian of the leads is
\begin{equation}
  \label{eq:H-leads-1}
  H_{\mlead} = \sum_{l=s,d}\sum_{\ve{k}\sigma} (\varepsilon_{l\ve{k}} - eV_l)
  \cre{c}{l\ve{k}\sigma} \ann{c}{l\ve{k}\sigma},
\end{equation}
where $e$ is the elementary charge and $V_l$ is the voltage in the
lead $l$. The operators $\cre{c}{l\ve{k}\sigma}$ and $\ann{c}{l\ve{k}\sigma}$ are
the creation 
and annihilation operators of a quasi-particle with wave vector
$\ve{k}$ and spin $\sigma =\pm$ in the lead $l$. 
The Hamiltonian of the gate is
\begin{equation*}
  H_g = -e \mu_g \mN,
\end{equation*}
where $\mu_g$ is the chemical potential in the gate and the operator
$\mN$ accounts for  the
total number in the QD system.
$H_T$ is the tunneling Hamiltonian
describing the tunneling between the QD and the two leads and it has
the form 
\begin{equation}
  \label{eq:H-tunnel}
  H_T = \sum_{l=s,d}\sum_{\beta\sigma} \int d\ve{r} \;  T_{l\beta}(\ve{r})
  \cre{\Psi}{\beta\sigma}(\ve{r})\Phi_{l\sigma}(\ve{r}) + \mathrm{H.c.},
\end{equation}
where $\Phi_{l\sigma}(\ve{r}) = \sum_{\ve{k}} \phi_{\ve{k}}(\ve{r})
\ann{c}{\ve{k}l\sigma}$ is the electron annihilation operator in the lead
$l$ and $\cre{\Psi}{\beta\sigma}(\ve{r})$ is the electron operator in the shell $\beta$ whose
explicit form will be given in Sec.~\ref{sec:low-energy-H}.

\subsection{Low energy non-interacting Hamiltonian of DWCNT}
\label{sec:low-energy-H}

\begin{figure}
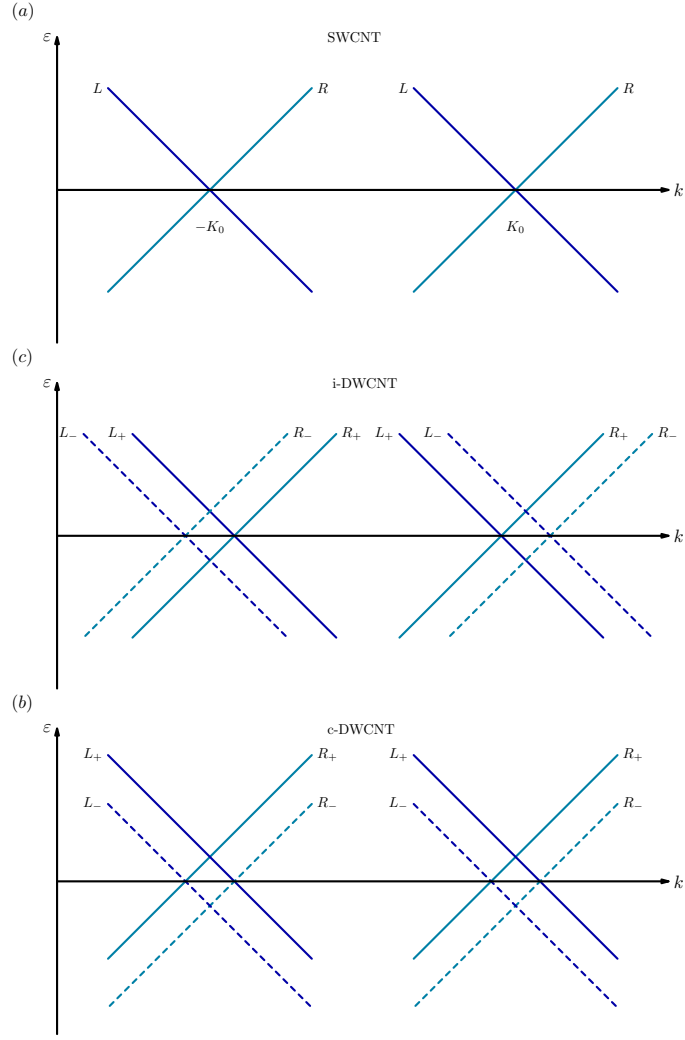

  \begin{center}
    \includegraphics[width=0.50\textwidth]{fig2a.eps} \qquad
    \includegraphics[width=0.50\textwidth]{fig2b.eps} \qquad
    \includegraphics[width=0.50\textwidth]{fig2c.eps} 
    
  \end{center}
  \caption{ (Color online) 
Energy spectra of metallic single-walled (SWCNT) and double-walled
(DWCNT) carbon nanotubes with \emph{periodic} boundary conditions
(PBC).
(a) 
Energy spectrum of a metallic SWCNT. There are two Fermi
points and 
two branches $L/R$ with left/right-moving electrons at
each the Fermi point. 
(b)
Energy spectrum 
of an incommensurate DWCNT (i-DWCNT). It consists of the energy
spectra of the outer and inner graphene shells ($\pm$), which are
not coupled to each other because of the
vanishing intershell coupling, cf. Eq.~\eqref{eq:H-linear-incom}.
(c) Energy spectrum
of a commensurate DWCNT (c-DWCNT). 
Because of the finite intershell coupling,
it is composed of the bonding and
anti-bonding bands ($\pm$), which are shift vertically along the
$\varepsilon$-axis, cf. Eq.~\eqref{eq:ham-com-linear}.
}
  \label{fig:energy-pbc}
\end{figure}

\begin{figure}[htbp]
  \includegraphics[width=0.3\textwidth]{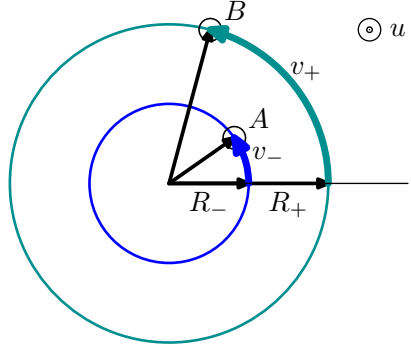}
  \caption{
(Color online)
Cross section of a DWCNT. Atoms $A$ and $B$ in two shells of
    radii $R_+$ and $R_-$, respectively, are projected onto this cross
    section. Such atoms are described by the coordinates $(u_+, v_+)$
    and $(u_-, v_-)$, where $u_{\pm}$ are along the tube axis and $v_{\pm}$
    measure the atom positions on the outer/inner circumference.
}
  \label{fig:graphene}
\end{figure}

\begin{figure}
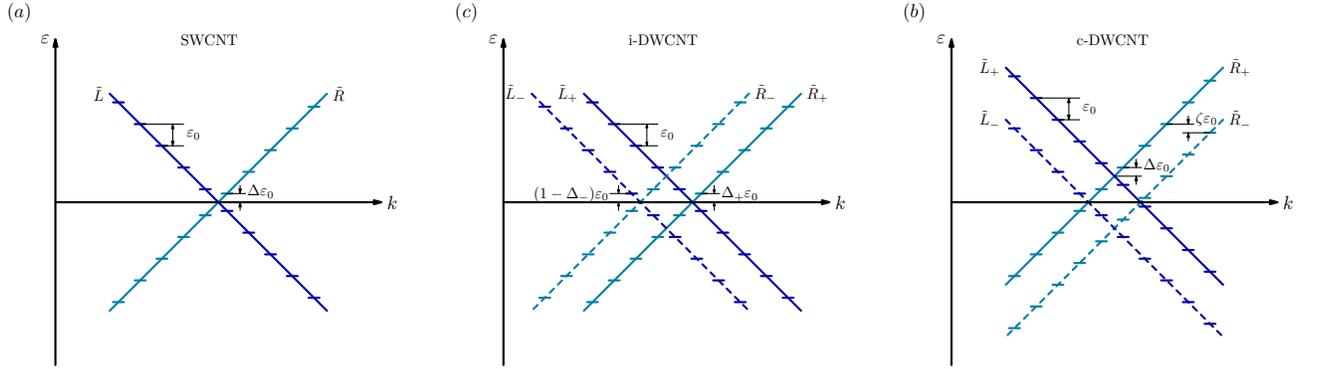

  \begin{center}
    \includegraphics[width=0.29\textwidth]{fig4a.eps} \qquad
    \includegraphics[width=0.29\textwidth]{fig4b.eps} \qquad
    \includegraphics[width=0.29\textwidth]{fig4c.eps} 
    
  \end{center}
  \caption{
(Color online) Energy spectra of metallic single-walled (SWCNT) and
double-walled (DWCNT) carbon nanotubes with \emph{open} boundary
conditions (OBC).
(a)
Energy spectrum of a metallic SWCNT. There are 
two branches $\Lt/\Rt$ with left/right-moving electrons.
The parameter $\varepsilon_0$ is the level spacing and $\Delta$
describes the 
mismatch of the Fermi point, cf. Eq.~\eqref{eq:k-quantization}.  (b)
Energy spectrum of a commensurate DWCNT
(c-DWCNT) and (c) of an incommensurate DWCNT (i-DWCNT).
The parameter $\Delta_{\pm}$ describes the mismatch of the Fermi point in the
shell $\pm$ and for a c-DWCNT is $\Delta_+=\Delta_-=\Delta$. The parameter $\zeta$
describes the mismatch of the states in two bands,
cf. Eq.~\eqref{eq:zeta}.
}
  \label{fig:energy-K}
\end{figure}

In general, the energy spectrum of a SWCNT or of a DWCNT without
electron-electron 
interactions can be obtained by using a
tight-binding model for the $p_z$ orbitals  in carbon
atoms~\cite{saito:1998}. In particular, we shall view in the following
a DWCNT as two tunneling coupled SWCNT shells. We denote with the
index $\beta=\pm$ the outer/inner SWCNT shell.
A metallic
SWCNT shell $\beta$ within periodic boundary conditions has two
independent Fermi points ($\pm K_{0,\beta}$). Their positions depend on the
chirality of the shell and in general are different for different SWCNT shells.
At the Fermi points, the lowest
conduction and the highest
valence bands touch each other as shown in
Fig.~\ref{fig:energy-pbc}(a). As the next 
conduction and valence 
bands are 
separated by a large gap (about $\unit[1]{eV}$)~\cite{saito:1998}, we
will only consider the 
lowest conduction band and the highest valence band in our
calculations. The energy
dispersion near the Fermi points is
linear~\cite{saito:1998}, see Fig.~\ref{fig:energy-pbc}(a), and is
given by  
\begin{equation}
  \label{eq:E-linear}
  \varepsilon_{R/L}(\kappa) = \pm \hbar v_F \kappa,
\end{equation}
where the wave vector $\kappa$ is measured
with respect to the Fermi points
and the Fermi velocity in SWCNTs is $v_F \approx \unit[8 \times 10^5]{m/s}$. 
Hence, at each Fermi point, there are two branches
corresponding to the right ($+$) and left ($-$) moving
electrons. The Bloch waves 
for the electrons in these branches in a shell $\beta=\pm$ are
\begin{equation}
  \label{eq:bloch}
  \varphi_{\beta r F \kappa}(\ve{r}) = e^{i\kappa u} \varphi_{\beta r F}(\ve{r}),
\end{equation}
where $\ve{r}=(u,v)$ and $u$ and $v$ are along the nanotube axis
and the circumference directions, respectively
(cf. Fig.~\ref{fig:graphene}).
The index $r=R/L=\pm$ denotes the right and
left-moving electrons. The periodic function is
\begin{equation}
  \label{eq:periodic-f}
  \varphi_{\beta rF} (\ve{r}) = \frac{1}{\sqrt{N_{\beta}}} \sum_{\ve{R}p} e^{i\ve{F}\cdot \ve{R}}
  f_{\beta prF} \, \chi(\ve{r}-\ve{R}-\veg{\tau}_p),
\end{equation}
where 
the index $\ve{R}$ denotes the lattice vector of the graphene sheet, $N_{\beta}$ is
the number of carbon atoms in the shell $\beta$, $p$ is the index
for the two graphene sublattices and $\veg{\tau}_p$ is the 
vector giving the positions of the two different atoms in a unit
cell. 
The index $F$ is for the Fermi points in the shell and $\ve{F}$
denotes
the Fermi points in a graphene sheet. They are related as
$F \equiv \ve{F}\cdot\ve{u} $ with $\ve{u}$ the unit
vector along the nanotube axis.
The
coefficients $f$'s depend on the chirality of the shell $(m,n)$
as~\cite{saito:1998}
\begin{align}
\label{eq:coeff-1}
  f_{\beta ArF} &= \frac{1}{\sqrt{2}\mL} \Bigl(- \frac{\sqrt{3}}{2}
  \sgn(F r)
    (n+m) + \frac{i}{2} \sgn(r) (m-n) \Bigr), \\
\label{eq:coeff-2}
  f_{\beta BrF} &= \frac{1}{\sqrt{2}}.
\end{align}
where
$\mL=\sqrt{n^2+mn+m^2}$ and 
the function $\chi(\ve{r} - \ve{R})$ is the $p_z$ orbital wave
function. 
Because we consider a finite length shell, we have to use the
open boundary 
condition (OBC) instead of the periodic boundary condition along the
tube axis (cf. Fig.~\ref{fig:energy-K}). The wave function in the shell
$\beta$ satisfying the OBC has 
the form~\cite{mayrhofer:epjb2007}
\begin{equation}
  \label{eq:wave-function-OBC}
  \varphi_{\beta \tilde{R}/\tilde{L} \kappa}^{OBC}(\ve{r}) = \frac{1}{\sqrt{2}} \Bigl(
  \varphi_{\beta \, R/L \, K_{0} \kappa}(\ve{r}) - \varphi_{\beta \, L/R \, -K_{0} \, -\kappa}(\ve{r}) \Bigr),
\end{equation}
and the wave vectors $\kappa$ are quantized as
\begin{equation}
  \label{eq:k-quantization}
  \kappa = \frac{\pi}{L} ( m_\kappa + \Delta_{\beta}), \qquad m_\kappa = 0, \pm1, \pm 2, \cdots,
\end{equation}
where $L$ is the length of the nanotube. 
The mismatch of the Fermi points is
$0 \leq \Delta_{\beta} = K_{0,\beta} L/ \pi -
[K_{0,\beta}L/ \pi] < 1$, where
$[\cdots]$ gives the integer part of its argument.
 The Hamiltonian of a finite length non-interacting shell $\beta$ is thus
\begin{equation}
  \label{eq:energy}
  H_{\beta}^0= \sum_{\rt\sigma \kappa} \sgn(\rt ) \hbar v_F \kappa \,  \cre{c}{\beta\rt
    \kappa \sigma}
  \ann{c}{\beta\rt \kappa \sigma},
\end{equation}
where $\rt =  \Rt/ \Lt = \pm $ is the index for the left and right moving
electrons with the OBC. 
 The operators $\cre{c}{\beta \rt \kappa \sigma}$ and $\ann{c}{\beta \rt \kappa \sigma}$
are the creation and annihilation operators of an electron in the
branch $\rt$ with the wave vector $\kappa$ and spin $\sigma$.

Let us now see how the spectrum gets modified when looking at
DWCNT. At low energies, the intershell couplings are different for 
c-DWCNTs and i-DWCNTs. The intershell couplings in i-DWCNTs are
negligible while they are quite strong in
c-DWCNTs~\cite{saito:jap1993, roche:prb2001, wang:2005, 
 yoon:prb2002, uryu:prb2004, uryu:245403}. 
Therefore, the non-interacting Hamiltonian of an i-DWCNT is the
combination of the 
Hamiltonians of two shells,
\begin{equation}
  \label{eq:H-linear-incom}
  \begin{split}
  H_{\idwnt}^{0} &= \sum_{\beta \rt \sigma \kappa} \sgn(\rt)\hbar v_F  \kappa \,
\cre{c}{\beta\rt \kappa \sigma} \ann{c}{\beta\rt \kappa \sigma} \\
  &= \sum_{\beta \rt \sigma m_\kappa } \sgn(\rt) (m_\kappa \varepsilon_0 + \Delta_\beta\varepsilon_0) 
\cre{c}{\beta\rt \kappa \sigma} \ann{c}{\beta\rt \kappa \sigma},
  \end{split}
\end{equation}
where $\varepsilon_0 = \hbar v_F \pi/L$ is the level spacing and we have used the
quantization relation for $\kappa$,
Eq.~\eqref{eq:k-quantization} (cf. Fig.~\ref{fig:energy-K}(b)).
On the other hand, the non-interacting Hamiltonian of a c-DWCNT
contains also the 
contribution from the intershell coupling~\cite{wang:2005},
\begin{equation}
  \label{eq:H-commensurate}
  H_{\cdwnt}^{0} =  \sum_{\beta}   \sum_{\rt\sigma \kappa} \sgn(\rt) \hbar v_F\kappa \,
  \cre{c}{\beta\rt \kappa \sigma } \ann{c}{\beta\rt \kappa \sigma } + 
  \sum_{\beta\beta^\p} \sum_{\rt  \sigma \kappa} t \cre{c}{\beta\rt \kappa \sigma } \ann{c}{\beta^\p \rt  \kappa \sigma} +
  \mathrm{H.c.},
\end{equation}
where $t$ is the intershell coupling and we assume that it is a
constant in the low energy regime. The Hamiltonian
Eq.~\eqref{eq:H-commensurate} can
be diagonalized by using the bonding and anti-bonding basis,
\begin{equation}
\label{eq:transformation}
 \ann{\tc}{\nu \rt \kappa \sigma} = \frac{1}{\sqrt{2}} (\ann{c}{+ \rt \kappa \sigma} +
 \sgn(\nu) \ann{c}{- \rt \kappa \sigma}), \qquad  
 \cre{\tc}{\nu \rt \kappa \sigma} = \frac{1}{\sqrt{2}} (\cre{c}{+ \rt \kappa
   \sigma } +
 \sgn(\nu) \cre{c}{- \rt \kappa \sigma}),
\end{equation}
where $\nu = \pm$ is the index for bonding and anti-bonding states,
respectively.
The non-interacting Hamiltonian of a c-DWCNT in the new basis
becomes 
\begin{equation}
  \label{eq:ham-com-linear}
  \begin{split}
  H_{\cdwnt}^{0} &= \sum_{\nu \rt \sigma \kappa}
  ( \sgn(\rt) \hbar v_F  \kappa + \sgn(\nu) t) 
\cre{\tc}{\rt \nu \kappa \sigma}\ann{\tc}{\rt \nu \kappa \sigma}  \\
  & = \sum_{\nu \rt \sigma m_\kappa} (\sgn(\rt)(m_\kappa \varepsilon_0 + \Delta\varepsilon_0) + \sgn(\nu)\zeta\varepsilon_0)
\cre{\tc}{\rt \nu \kappa \sigma}\ann{\tc}{\rt \nu \kappa \sigma}.
  \end{split}
\end{equation}
where for c-DWCNT is $\Delta_\beta=\Delta$ and the parameter $\zeta$ is defined as
\begin{equation}
  \label{eq:zeta}
0 \leq \zeta = t/ \varepsilon_0 - [t/ \varepsilon_0] < 1,
\end{equation}
which describes the mismatch of states in two bands
(cf. Fig.~\ref{fig:energy-K}(c)). 

\subsection{Coulomb interaction Hamiltonian of DWCNTs}
\label{sec:H-Coulomb}

In quasi-one dimensional electronic structures as CNTs, Coulomb
interactions are not fully screened and can strongly influence 
the properties of CNTs~\cite{bockrath:nature1999,postma2001cns,
  kane:prl1997, egger:prl1997, egger:epjb1998, egger:prl1999, wang:2005}. 
The total Coulomb interactions in an i-DWCNT can
be expressed by the following Hamiltonian,
\begin{equation}
  \label{eq:Coulomb-interaction}
  H^{int}_{\idwnt} = \frac{1}{2} \sum_{\beta\beta^\p\sigma\sigma^\p} \iint d\ve{r}_1 d\ve{r}_2
  \cre{\Psi}{\beta\sigma}(\ve{r}_1)\cre{\Psi}{\beta^\p\sigma^\p}(\ve{r}_2) 
  U_{\beta\beta^\p}(\ve{r}_1 - \ve{r}_2)
 \ann{\Psi}{\beta^\p\sigma^\p}(\ve{r}_2) \ann{\Psi}{\beta\sigma}(\ve{r}_1) ,
\end{equation}
where $\ve{r}_i=(u_i, v_i)$ and $u$ and $v$ are along the tube axis
and the circumference direction, respectively
(cf. Fig.~\ref{fig:graphene}) .  
The intrashell interaction is given as
\begin{equation}
  \label{eq:interaction-kernel}
  U_{\beta\beta}(\ve{r}_1 - \ve{r}_2) = \frac{e^2 / \epsilon }{\sqrt{(u_1 - u_2)^2 + 4
      R_{\beta}^2 \sin^2 \bigl( (v_1-v_2)/2R_\beta) + a_z^2}},
\end{equation}
and the intershell interaction is
\begin{equation}
  \label{eq:interaction-kernel-1}
  U_{+ -}(\ve{r}_1 - \ve{r}_2) = \frac{e^2 / \epsilon }{\sqrt{(u_1 - u_2)^2 + 4
      R_{+}R_- \sin^2 ( v_1/2R_+-v_2/2R_- ) + \Delta R^2}},
\end{equation}
where $\epsilon$ is the dielectric constant, $a_z$ is the
``thickness'' of a graphene sheet and the distance between two shells is
$\Delta R = \abso{R_+ - R_-}$. 
The electron operators for i-DWCNTs are defined as
\begin{equation}
  \label{eq:e-op}
 \Psi_{\beta \sigma}(\ve{r}) \equiv
   \sum_{\rt q} \varphi_{\beta \rt q}^{OBC}(\ve{r}) \ann{c}{\beta \rt \sigma q} 
\end{equation}
Using Eq.~\eqref{eq:bloch} we can define the 1D electron operators 
describing the slowly varying 
part of the electron operators $\Psi_{\beta\sigma}(\ve{r})$ as
\begin{equation}
  \label{eq:e-op-1d}
  \psi_{\beta \rt F \sigma }(u) = \frac{1}{\sqrt{2L}} \sum_{q} e^{i\sgn(F)qu}
  \ann{c}{\beta \rt \sigma q},
\end{equation}
in terms of which the electron operators can be written as
\begin{equation}
  \label{eq:e-op-idwcnt}
  \begin{split}
  \Psi_{\beta\sigma}(\ve{r}) &= \sum_{\rt q}\varphi_{\beta \rt q}^{OBC}(\ve{r}) \ann{c}{\beta
    \rt \sigma
    q} \\
  &= \sqrt{L} \sum_{\rt F} \sgn(F) \varphi_{\beta\sgn(F) \rt  F}(\ve{r})
  \psi_{\beta \rt F\sigma }(u).
  \end{split}
\end{equation}
For SWCNT shells with diameter larger than $\sim \unit[1]{nm}$, we can
only keep forward-scattering (or density-density)
processes~\cite{mayrhofer}, such that
the interacting Hamiltonian in an i-DWCNT becomes 
\begin{equation}
  \label{eq:H-interaction-1D}
  H_{\idwnt}^{int} = \frac{1}{2} \sum_{\beta\beta^\p}\sum_{\rt\rt^\p}\sum_{FF^\p}\sum_{\sigma\sigma^\p} \iint
  du_1 du_2 \, \rho_{\beta\rt F\sigma}(u_1) V^{eff}_{\beta\beta^\p} (u_1-u_2) \rho_{\beta^\p\rt^\p F^\p
    \sigma^\p}(u_2),
\end{equation}
where $\rho_{\beta\rt F\sigma} (u) = \cre{\psi}{\beta \rt F\sigma}(u) \ann{\psi}{\beta \rt F\sigma}(u)$ is the
electron density operator and $V^{eff}_{\beta\beta^\p}$ is the effective 
one-dimensional Coulomb interactions given by
\begin{equation}
  \label{eq:V-eff}
  V^{eff}_{\beta\beta^\p} (u_1, u_2) = \frac{L^2}{N_{\beta} N_{\beta^\p}} \sum_{\ve{R}_1\ve{R}_2} \iint
  dv_1 dv_2 \, \abso{\chi(\ve{r}_1 - \ve{R}_1)}^2 U_{\beta\beta^\p}(\ve{r}_1 -
  \ve{r}_2) \abso{\chi(\ve{r}_2 - \ve{R}_2)}^2 .
\end{equation}

Let us now turn to c-DWCNTs. The total Coulomb interaction Hamiltonian
of a c-DWCNT has a similar form as 
Eq.~\eqref{eq:Coulomb-interaction}. However, we have to rewrite it in
the basis of
the bonding/anti-bonding states using the transformation 
Eq.~\eqref{eq:transformation}. In this basis, the Coulomb interaction
of a c-DWCNT is given by
\begin{equation}
  \label{eq:Coulomb-interaction-c}
  H_{\cdwnt}^{int} = \frac{1}{2} \sum_{\nu\nu^\p\sigma\sigma^\p} \iint d\ve{r}_1 d\ve{r}_2
  \cre{\Psi}{\nu\sigma}(\ve{r}_1)\cre{\Psi}{\nu^\p\sigma^\p}(\ve{r}_2) 
  \tU_{\nu\nu^\p}(\ve{r}_1 - \ve{r}_2)
  \ann{\Psi}{\nu^\p\sigma^\p}(\ve{r}_2) \ann{\Psi}{\nu\sigma}(\ve{r}_1),
\end{equation}
where $\nu=\pm$ is the index for bonding and anti-bonding bands and the
new interactions are 
\begin{equation}
 \label{eq:U-com}
 \tU_{+ -} = 2\tU_{++} = 2\tU_{--} = \frac{1}{4} (U_{++} + U_{--} +
 U_{+ -}),
\end{equation}
where $U_{\beta\beta^\p}$ is defined by Eqs.~\eqref{eq:interaction-kernel}
and~\eqref{eq:interaction-kernel-1}. The electron operators in
c-DWCNTs are defined as 
\begin{equation}
  \label{eq:e-op-cdwcnt}
  \begin{split}
  \Psi_{\nu \sigma}(\ve{r}) &= \sum_{\rt q} \tilde{\varphi}_{\nu \rt q}^{OBC}(\ve{r})
  \ann{\tc}{\nu \rt \sigma
    q} \\
  &= \sqrt{L} \sum_{\rt F} \sgn(F) \tilde{\varphi}_{\nu \sgn(F) \rt F}(\ve{r})
  \tilde{\psi}_{\nu \rt\sigma F}(u),
  \end{split}
\end{equation}
where $\tilde{\varphi}_{\nu\rt q}^{OBC}$ is a linear combination of $\varphi_{\beta\rt
  q}^{OBC}$, namely, $\tilde{\varphi}_{\nu \rt q}^{OBC}(\ve{r}) = \bigl((\varphi_{+ \rt
  q}^{OBC}(\ve{r}) +
\sgn(\nu) \varphi_{- \rt q}^{OBC}(\ve{r}) \bigr)/ \sqrt{2}$.
By using these electron operators and keeping only the relevant
forward-scattering processes, the Coulomb interaction Hamiltonian
of a c-DWCNT becomes
\begin{equation}
  \label{eq:H-interaction-c}
  H_{\cdwnt}^{int} = \frac{1}{2} \sum_{\nu\nu^\p}\sum_{\rt\rt^\p}\sum_{FF^\p}\sum_{\sigma\sigma^\p} \iint
  du_1 du_2 \, \tilde{\rho}_{\nu\rt F\sigma}(u_1) \tilde{V}^{eff}_{\nu\nu^\p}
  (u_1-u_2) \tilde{\rho}_{\nu^\p\rt^\p F^\p
    \sigma^\p}(u_2),
\end{equation}
where $\tilde{\rho}_{\nu\rt F\sigma} (u) = \cre{\tilde{\psi}}{\nu \rt
  F\sigma}(u) \ann{\tilde{\psi}}{\nu \rt F\sigma}(u)$ is the
density operator and $\tilde{V}^{eff}_{\nu\nu^\p}$ is the effective 
one-dimensional Coulomb interaction,
\begin{equation}
  \label{eq:V-eff-c}
  \tilde{V}^{eff}_{\nu\nu^\p} (u_1, u_2) = \frac{L^2}{N_{\nu}N_{\nu^\p}}
  \sum_{\ve{R}_1\ve{R}_2} \iint 
  dv_1 dv_2 \, \abso{\chi(\ve{r}_1 - \ve{R}_1)}^2 \tU_{\nu\nu}(\ve{r}_1 -
  \ve{r}_2) \abso{\chi(\ve{r}_2 - \ve{R}_2)}^2 .
\end{equation}

\subsection{Bosonization}
\label{sec:bosonization}

The low energy Hamiltonian of a DWCNT is the combination of the
non-interacting and interacting Hamiltonians and it can be 
diagonalized by the bosonization method~\cite{haldane1981llt,
  voit:rpp1994,delft:adp1998,giamarchi:2004}. First, we
introduce the bosonic operators~\cite{delft:adp1998, wang:2005,
  mayrhofer:epjb2007}
\begin{equation}
  \label{eq:boson-b}
  b_{\alpha \sgn(\rt) q \sigma } = \begin{cases}
\rho_{\alpha \rt q \sigma} / \sqrt{n_q}, & \text{for i-DWCNTs}, \\
\tilde{\rho}_{\alpha \rt q \sigma}/ \sqrt{n_q}, & \text{for c-DWCNTs}, 
  \end{cases}
\end{equation}
where $q = \pi n_q/L$ with $n_q$ an integer. 
The index $\alpha=\pm$
denotes the bonding/anti-bonding states in c-DWCNTs and outer/inner
shells in i-DWCNTs and we will keep this convention in the rest of
the paper. The bosonic
operators obey the bosonic commutation relation
\begin{equation*}
  \com{\ann{b}{\alpha q \sigma} , \cre{b}{\alpha^\p q^\p \sigma^\p}} =
  \delta_{\alpha\alpha^\p}\delta_{qq^\p}\delta_{\sigma\sigma^\p}.
\end{equation*}
By using these bosonic operators, the Hamiltonian of a DWCNT QD can be
separated into its fermionic and bosonic parts, $H_{QD} = H_f +
H_{b}$.  The fermionic part $H_f$
describes the ground state and the fermionic excitations in the
DWCNT. The fermionic Hamiltonian of a c-DWCNT is
\begin{equation}
  \label{eq:H-groundstate-cdwcnt}
  H_{f,\cdwnt} = 
   \sum_{\alpha\rt \sigma} \frac{1}{2} \varepsilon_0 \mN_{\alpha\rt \sigma}^2 + \Delta\varepsilon_0
   \sgn(\rt) \mN_{\alpha \rt
    \sigma} + \Bigl(\alpha \zeta \varepsilon_0 - \frac{1}{2}\varepsilon_0 \Bigr) \mN_{\alpha \rt \sigma}
  +H_f^{int},
\end{equation}
while the i-DWCNT fermionic Hamiltonian has the form
\begin{equation}
  \label{eq:H-groundstate-idwcnt}
  H_{f,\idwnt} = \sum_{\alpha \rt \sigma} \frac{1}{2} \varepsilon_0 \mN_{\alpha \rt \sigma}^2 + \Delta_\alpha \varepsilon_0 \sgn(\rt)
  \mN_{\alpha \rt \sigma}   - \frac{1}{2}\varepsilon_0 \mN_{\alpha  \rt \sigma}  + H_f^{int},
\end{equation}
where $H_f^{int}$ is due to Coulomb interaction having the form 
\begin{equation}
  \label{eq:H0I}
  H_f^{int} = \frac{1}{2}  \sum_{\alpha\alpha^\p} W_{00}^{\alpha\alpha^\p} \Bigl(\sum_{\rt\sigma}
  \mN_{\alpha\rt\sigma}\Bigr) \Bigl(\sum_{\rt^\p\sigma^\p} \mN_{\alpha^\p\rt^\p\sigma^\p} \Bigr),
\end{equation}
with the interaction strengths
\begin{equation*}
  W_{00}^{\alpha\alpha^\p} = \frac{1}{L^2}\iint du_1du_2 \, V_{\alpha\alpha^\p}^{eff}(u_1 -
  u_2).
\end{equation*}
Therefore, the fermionic Hamiltonian of a DWCNT QD is described by the
constant-interaction model~\cite{grabert1991set}. 

The bosonic excitations of a DWCNT QD are described by the Hamiltonian
$H_{b}$, which can be expressed in terms of the bosonic operators as
\begin{equation}
  \label{eq:H-ex}
  \begin{split}
H_{b} &= \sum_{q \neq  0} \sum_{\alpha\sigma \rt} \varepsilon_0 \abso{n_q} \cre{b}{\alpha \sgn(\rt) q \sigma} \ann{b}{\alpha \sgn(\rt)
    q \sigma} \\
  &\qquad + \frac{1}{2}  \sum_{q>0} \sum_{\alpha\alpha^\p\rt\rt^\p \sigma\sigma^\p} n_q
W_{qq}^{\alpha\alpha^\p} \bigl(\ann{b}{\alpha \sgn(\rt)q \sigma} + \cre{b}{\alpha \sgn(\rt) q
  \sigma} \bigr)
 \bigl( \ann{b}{\alpha^\p \sgn(\rt^\p)q \sigma^\p} + \cre{b}{\alpha^\p \sgn(\rt^\p) q
   \sigma^\p} \bigr),
  \end{split}
\end{equation}
with the interaction strengths
\begin{equation}
  \label{eq:interaction-strength}
  W_{qq}^{\alpha\alpha^\p} = \frac{1}{L^2}\iint du_1du_2 \,
  V_{\alpha\alpha^\p}^{eff}(u_1-u_2) \cos(qu_1) \cos(q u_2).
\end{equation}
In order to diagonalize the Hamiltonian $H_{b}$, we need to introduce
new bosonic operators $\ann{a}{j\delta\xi q}$'s, where $j=c,s$ denote
charge/spin modes and the remaining indices 
 $\delta=\pm$ and
$\xi=\pm$ define total/relative modes with respect to the branch and
shell (or bonding/anti-bonding state) degrees of freedoms,
respectively. The new bosonic operators are 
related to the bosonic operators $\ann{b}{\alpha \sgn(\rt)q \sigma}$'s
as~\cite{matveev:prb1993, wang:2005}
\begin{equation}
  \label{eq:new-bosonic-op}
  \ann{b}{\alpha \sgn(\rt) q \sigma } = \sum_{j\delta\xi} \Lambda_{\alpha \rt \sigma}^{j\delta\xi q} ( S_{j\delta\xi q}
  \ann{a}{j\delta\xi q} + C_{j\delta\xi q} \cre{a}{j\delta\xi q} ),
\end{equation}
where the coefficients are given by
\begin{equation}
  \label{eq:Lambda}
  \Lambda_{\alpha\rt\sigma}^{j\delta \xi q} = \frac{1}{2\sqrt{2}}
  \begin{pmatrix}
    \sin\theta_q + \cos\theta_q & - \cos\theta_q + \sin\theta_q & 1 & 1 & 1 &1 & 1& 1 \\
    \sin\theta_q + \cos\theta_q & - \cos\theta_q + \sin\theta_q & 1 & 1 & -1 & -1 & -1 & -1 \\
    \sin\theta_q + \cos\theta_q & - \cos\theta_q + \sin\theta_q & -1 & -1 & 1 & 1 & -1 & -1 \\
    \sin\theta_q + \cos\theta_q & - \cos\theta_q + \sin\theta_q & -1 & -1 & -1 & -1 & 1 & 1 \\
    \sin\theta_q + \cos\theta_q & - \cos\theta_q - \sin\theta_q & 1 & -1 & 1 & -1 & 1 & -1 \\
    \sin\theta_q + \cos\theta_q & - \cos\theta_q - \sin\theta_q & 1 & -1 & -1 & 1 & -1 & 1 \\
    \sin\theta_q + \cos\theta_q & - \cos\theta_q - \sin\theta_q & -1 & 1 & 1 & -1 & -1 & 1 \\
    \sin\theta_q + \cos\theta_q & - \cos\theta_q - \sin\theta_q & -1 & 1 & -1 & 1 & 1 & -1 
  \end{pmatrix},
\end{equation}
with 
\begin{equation*}
  \sin\theta_q = \Bigl\lvert W_{qq}^{++} - W_{qq}^{--} \Bigr\rvert \Biggl/ \Biggl(
  \Bigl(W_{qq}^{++} - W_{qq}^{--} \Bigr)^2 + \Bigl( W_{qq}^{+ -}
   +
  \sqrt{\Bigl(W_{qq}^{++} - W_{qq}^{--} \Bigr)^2 + \Bigl( W_{qq}^{+ -}
    \Bigr)^2} \, \Bigr)^2 \Biggr)^{1/2}.
\end{equation*}
The other coefficients are
\begin{equation}
\label{eq:SC1}
S_{j\delta\xi q} = 1 \quad  \text{and } \quad C_{j\delta\xi q}=0
\end{equation}
in the cases $(j\delta\xi) =(c-\pm), (s\pm\pm)$.
For the total and relative 
charge modes ($c+\pm$), the two coefficients are interaction dependent
\begin{equation}
  \label{eq:SC}
S_{c+\pm q} = \frac{1}{2} \Biggl( \sqrt{\frac{\varepsilon_{0}}{\varepsilon_{c+\pm}(q)}} +
\sqrt{\frac{\varepsilon_{c+\pm}(q)}{\varepsilon_{0}}} \Biggr), \qquad
C_{c+\pm q} = \frac{1}{2} \Biggl( \sqrt{\frac{\varepsilon_{0}}{\varepsilon_{c+\pm}(q)}} -
\sqrt{\frac{\varepsilon_{c+\pm}(q)}{\varepsilon_{0}}} \Biggr),
\end{equation}
where the energies of the total and relative charge modes are
\begin{equation}
  \label{eq:energy-charge}
\varepsilon_{c+\pm}(q) = \varepsilon_0\sqrt{1 + 8 W_{qq}^{\pm\pm}/ \varepsilon_0} \;.
\end{equation}
%
%
The interactions do not affect the 6 ``neutral'' modes, $(j\delta\xi) =
(c-\pm), (s\pm\pm)$ and
their energy dispersions are the same as for the non-interacting
system, 
\begin{equation}
  \label{eq:energy-neutral}
  \varepsilon_{j\delta\xi}(q) = \varepsilon_0. 
\end{equation}
By using the
new bosonic operators, the excitation Hamiltonian can be diagonalized
to be
\begin{equation}
  \label{eq:H-ex-diagonalized}
  H_{b} = \sum_{q>0}\sum_{j\delta\xi} \varepsilon_{j\delta\xi}(q) \cre{a}{j\delta\xi q} \ann{a}{j\delta\xi q},
\end{equation}
and the eigenstates are
\begin{equation}
  \label{eq:Hex-states}
  \ket{\ve{N}, \ve{m}} \equiv \prod_{q>0} \prod_{j\delta\xi} \frac{1}{\sqrt{m_{j\delta\xi q}!}}
  \Bigl(\cre{a}{j\delta\xi q} \Bigr)^{m_{j\delta\xi q}} \ket{\ve{N}, \boldsymbol{0}},
\end{equation}
where $\ve{N} = \{ N_{\alpha \rt\sigma} \}$ defines the number of electrons in each
of the eight branches $(\alpha \rt \sigma)$ and $\ve{m} = \{ m_{j\delta\xi q} \}$
describes the configuration of the bosonic excitations in
each of the eight modes $(j\delta \xi)$. The state $\ket{\ve{N}, \ve{0}}$
contains no bosonic excitations and describes the ground state or
the fermionic excited states.

\section{Dynamics of the QD system}
\label{sec:dynamics-qd}

The transport properties of the DWCNT QD system can be obtained by
investigating the dynamics of its density matrix~\cite{blum:1996}. In
this section, we briefly show how to derive the equation of motion for the
reduced density matrix of the DWCNT QD system. By solving these
equations we obtain
the stationary current through the DWCNT QD system when a bias
voltage is applied. 

\subsection{Equation of motion for the reduced density matrix}
\label{sec:density-matrix}

As we consider a very weak coupling between the DWCNT QD and the two
leads, the tunneling Hamiltonian can
be treated as a perturbation and we can obtain the equation for motion
for the density matrix in the interaction picture as~\cite{blum:1996}
\begin{equation}
  \label{eq:eom-dm}
  i\hbar \frac{\partial \rho_{tot}^I(t)}{\partial t} = \com{H^I_T(t), \rho^I_{tot}(t)},
\end{equation}
where $\rho_{tot}^I(t)$ is the density matrix of the whole system (including
the DWCNT QD and the 
leads) and the tunneling Hamiltonian in the interaction
picture is 
\begin{equation}
  H^I_T(t) = e^{\frac{i}{\hbar} (H_{QD} + H_{\mlead} )(t-t_0)} H_T
  e^{-\frac{i}{\hbar} (H_{QD} + H_{\mlead} )(t-t_0)}.
\end{equation}
This equation can be solved formally as
\begin{equation}
  \rho_{tot}^I(t) = \rho_{tot}^I(t_0) - \frac{i}{\hbar} \int_{t_0}^{t} dt_1 \, \com{H_T^I(t_1),
    \rho_{tot}^I(t_1)}.
\end{equation}
Substituting the above expression of $\rho_{tot}^I(t)$ back to
Eq.~\eqref{eq:eom-dm}, we have
\begin{equation}
  \label{eq: eom-dm-1}
 \frac{\partial \rho_{tot}^I(t)}{\partial t} (t) = -\frac{i}{\hbar} \com{H_T^I(t), \rho_{tot}^I(t_0)}
 +  \Bigl(\frac{i}{\hbar}\Bigr)^2 \int_{t_0}^{t} dt_1 \, \com{H_T^I(t), \com{H_T^I(t_1),
    \rho_{tot}^I(t_1)}}.
\end{equation}
As we are only interested in the transport through the DWCNT QD, we
will focus on the reduced density matrix of the QD which is obtained
by tracing out the degrees of freedom of the leads,
\begin{equation}
  \label{eq:rho-rdm}
  \rho^I = \tro{\mlead}{\rho_{tot}^I}.
\end{equation}
Because the leads are very large comparing with the QD and
the tunneling events between leads and the QD are rare,
the effect of the QD on the leads can be ignored and 
the leads can be described as 
reservoirs remaining in thermal equilibrium. We use the 
ansatz~\cite{mayrhofer:epjb2007}  to factorize the total density
matrix $\rho^I(t)$,
\begin{equation}
  \label{eq:rho-factor}
  \rho_{tot}^I(t) = \rho^I_{\mlead}\rho^I(t) = \rho^I_{s}\rho^I_{d}\rho^I(t),
\end{equation}
where the density matrix of the leads, $\rho_{\mlead}$, is time
independent and is described by the thermal equilibrium distribution, 
\begin{equation*}
  \rho^I_{s/d} = \frac{e^{-\beta(H_{s/d}-\mu_{s/d}\mN_{s/d})}}
  {\tr{e^{-\beta(H_{s/d}-\mu_{s/d} \mN_{s/d})}}},
\end{equation*}
where $\mu_{s/d}$ is the chemical potential of the source/drain lead and
$\beta = 1/k_BT$.
We further simplify Eq.~\eqref{eq: eom-dm-1} by introducing the Markov
approximation, that is, we assume that $\dot{\rho}_{tot}^I(t)$ only locally
depends on $\rho_{tot}^I(t)$ and we can replace 
$\rho_{tot}^I(t^\p)$ by $\rho_{tot}^I(t)$.

We make the further assumptions that the elements of the reduced
density matrix 
between two states with different charges vanish, and that the elements
between two non-degenerate states with same charges also
vanish~\cite{mayrhofer:prb2006, mayrhofer:epjb2007}.
Finally, the master equations of the reduced density matrix
 can be expressed in Bloch-Redfield form~\cite{bloch:pr1957,redfield1957trp}
\begin{equation}
\label{eq:eom-rdm-BR}
\dot{\rho}_{nm}^{I,E_{N}}(t) = -\sum_{kk^\p}R_{nm
  kk^\p}^{E_{N}}\rho_{kk^\p}^{I,E_{N}}(t)
+\sum_{M=N\pm1}\sum_{E^\p}\sum_{kk^\p}R_{nm kk^\p}^{E_{N}\,
  E^\p_{M}}\rho_{kk^\p}^{I,E^\p_{M}}(t),
\end{equation}
where $n$,$m$,$k$, and $k^\p$ are indices of the eigenstates of the
DWCNT QD Hamiltonian. The Redfield tensors have the form
 \begin{align}
   \label{eq:Blochredf}
   R_{nm\, kk'}^{E_{N}} &= \sum_{l}\sum_{M,E^\p,j} \Bigl(\delta_{mk'}\Gamma_{l,njjk}^{(+)E_{N}\,
    E'_{M}}+\delta_{nk}\Gamma_{l,k^\p jjm}^{(-)E_{N}\, E'_{M}}\Bigr), \\
   R_{nm\, kk'}^{E_{N}\, E'_{M}}& =\sum_{l} \Gamma_{l \,k^\p mnk}^{(+)E'_{M}\,
     E_{N}} 
   + \Gamma_{l \,k^\p mnk}^{(-)E'_{M}\,
     E_{N}}, 
\end{align}
and the matrix elements of the electron operators are 
 \begin{equation*} 
\left(\Psi_{\alpha \sigma}^{\dagger}(\ve{x})\right)_{km}^{E_{N}\, E'_{N+1}} = \left\langle
  \ve{N}, 
  \ve{k} \left| \cre{\Psi}{\alpha \sigma}(\ve{x}) \right|\ve{N+1}, \ve{m} \right\rangle
\end{equation*}
with the states $\left|\ve{N}, \ve{k} \right\rangle $ and $\left|\ve{N+1},
  \ve{m}\right\rangle $
having energy $E_{N} $, $E'_{N+1}$ and particle number $N$, $N+1$,
respectively. Such matrix elements are calculated in analytic form in
App.~\ref{sec:matrix-e-op}. The transition rates depend on the
properties of the 
contacts between the leads and the
DWCNT QD~\cite{mayrhofer:epjb2007}. Here, we assume that the contacts
are  very simple. They do not mix the electrons in the different
branches and the couplings between the leads and the DWCNT do not
depend on either the wave vectors or the spins of the tunneling
electrons. Then the 
transition rates depend on the energy of the tunneling electrons
because of the matrix elements of the electron operators,
Eq.~\eqref{eq:element-1d-e-op} and have the forms (the derivation of
these expressions is given in App.~\ref{sec:tunneling-rate}), 
\begin{equation}
\label{eq:trans-rate-final-n+1}
  \begin{split}
    \Gamma_{l\, k^\p mnk}^{(\pm)E_{N}\, E_{N+1}} 
&= \sum_{\alpha} 
     \frac{\gamma_{l\alpha}}{h} g_l(\varepsilon_l) f(\varepsilon_l) \sum_{\rt \sigma F}
     \delta_{\ve{N} + \ve{e}_{\alpha\rt\sigma}, \, \ve{N+1}}
     \prod_{q >0} \prod_{q^\p>0} \prod_{j\delta\xi} \prod_{j^\p\delta^\p\xi^\p}  \\
&\qquad \times    F(\lambda_{\alpha\rt\sigma q}^{j\delta\xi F} (u_l), k_{j\delta\xi q}^{\p}, m_{j\delta\xi q})
    F^\ast(\lambda_{\alpha\rt\sigma q}^{j^\p\delta^\p\xi^\p F} (u_l), n_{j^\p\delta^\p\xi^\p
      q^\p}, k_{j^\p\delta^\p\xi^\p q^\p}) 
  \end{split}
\end{equation}
where the constants $\gamma_{l\alpha}$ describes the coupling strengths
between the bonding/anti-bonding state $\alpha$ in c-DWCNTs or between the
shell $\alpha$ in i-DWCNTs and the leads $l$ and $u_l=0,L$ for $l=s,d$. The
vector $\ve{e}_{\alpha\rt\sigma}$ 
denotes a state with one particle in the branch $\alpha\rt\sigma$.
The function $F(\lambda, m, m^\p)$ is defined in
Eq.~\eqref{eq:F-function} and the parameters $\lambda_{\alpha\rt \sigma q}^{j\delta\xi
  F}(u)$'s are defined in Eq.~\eqref{eq:lambda-2}.
The eigenstates involved are
\begin{align*}
 \ket{k^\p} &= \ket{\ve{N}, \ve{k}^\p},& \ket{m} &= \ket{\ve{N+1},
   \ve{m}}, \\
 \ket{n} &= \ket{\ve{N+1}, \ve{n}}, & \ket{k} &= \ket{\ve{N},
   \ve{k}}. 
\end{align*}
Similarly, the expressions for the remaining tunneling rates are
\begin{equation}
\label{eq:trans-rate-final-n-1}
  \begin{split}
    \Gamma_{l\, k^\p mnk}^{(\pm)E_{N}\, E_{N-1}} 
&= \sum_{\alpha} 
     \frac{\gamma_{l\alpha}}{h} g_l(\varepsilon_l) (1-f(\varepsilon_l)) \sum_{\rt \sigma F}
     \delta_{\ve{N} - \ve{e}_{\alpha\rt\sigma}, \, \ve{N-1}}
     \prod_{q >0} \prod_{q^\p>0} \prod_{j\delta\xi} \prod_{j^\p\delta^\p\xi^\p} \\
&\qquad \times    F^\ast(\lambda_{\alpha\rt\sigma q}^{j\delta\xi F} (u_l), k_{j\delta\xi q}^{\p}, m_{j\delta\xi q})
    F(\lambda_{\alpha\rt\sigma q}^{j^\p\delta^\p\xi^\p F} (u_l), n_{j^\p\delta^\p\xi^\p
      q^\p}, k_{j^\p\delta^\p\xi^\p q^\p}) 
  \end{split}
\end{equation}
with the eigenstates
\begin{align*}
 \ket{k^\p} &= \ket{\ve{N}, \ve{k}^{\p}},& \ket{m} &= \ket{\ve{N-1},
   \ve{m}}, \\
 \ket{n} &= \ket{\ve{N-1}, \ve{n}}, & \ket{k} &= \ket{\ve{N},
   \ve{k}}. 
\end{align*}
In the linear transport regime, only the following tunneling rates
between the 
ground states with $N$ and $N\pm1$ electrons are needed, which have very
simple expressions,
\begin{equation}
  \label{eq:rate-linear-n+1}
 \Gamma_{l\, knnk}^{(\pm) E_{N} \, E_{N+1}} = \sum_{\alpha} \sum_{\rt \sigma F}
 \frac{\gamma_{l\alpha}}{h} g_l(\varepsilon_l) f(\varepsilon_l) \;
     \delta_{\ve{N} + \ve{e}_{\alpha\rt\sigma}, \, \ve{N+1}}
\end{equation}
with the eigenstates $\ket{k} = \ket{\ve{N}, \ve{0}}$ and
$\ket{n} = \ket{\ve{N+1}, \ve{0}}$, and
\begin{equation}
  \label{eq:rate-linear-n-1}
 \Gamma_{l\, knnk}^{(\pm) E_{N} \, E_{N-1}} = \sum_{\alpha} \sum_{\rt \sigma F}
 \frac{\gamma_{l\alpha}}{h} g_l(\varepsilon_l) (1-f(\varepsilon_l)) \;
     \delta_{\ve{N} - \ve{e}_{\alpha\rt\sigma}, \, \ve{N-1}}
\end{equation}
with the eigenstates $\ket{k} = \ket{\ve{N}, \ve{0}}$ and
$\ket{n} = \ket{\ve{N-1}, \ve{0}}$.
We are only interested in the properties of the system
in the stationary state, which can be obtained by
 solving the
Eq.~\eqref{eq:eom-rdm-BR} with the left hand side set to be zero.

\subsection{Calculation of the current}
\label{sec:current}

The current can be calculated by using the
tunneling rates between the DWCNT QD and the leads. 
 The current measured in 
experiments is the current in one lead, which 
can be calculated as 
\begin{equation}
  \label{eq:current}
  I_l = e \sum_{N}\left(\Theta_{l}^{N\to N+1}-\Theta_{l}^{N\to N-1}\right),
\end{equation}
where $\Theta_l^{N \to N\pm 1}$ are the
tunneling rates between the QD and the 
lead $l$ when the particle number in the DWCNT QD changes from $N$ to $N \pm
1$. The tunneling rates are related to the transition
rates and the reduced density matrix as
\begin{equation}
  \label{eq:tunnel-rate}
  \Theta_{l}^{N\to N\pm1}= \sum_{E,E^\p}\sum_{nkj} \left(\Gamma_{l,njjk}^{(+)E_{N}\,
    E^\p_{N\pm1}}\rho_{kn}^{I,E_{N}} + \rho_{nk}^{I,E_{N}}\Gamma_{l,kjjn}^{(-)E_{N}\,
    E^\p_{N\pm1}} \right).
\end{equation}
After substituting Eq.~\eqref{eq:tunnel-rate} into
Eq.~\eqref{eq:current}, the current can be expressed in terms of the
transition rates and the elements of the reduced density matrix
as
\begin{equation}
  \label{eq:current-1}
  I_{l} = e \sum_{N,E,E'} 
  \left(\Gamma_{l,njjk}^{(+)E_{N}\, E^\p_{N+1}}-\Gamma_{l,njjk}^{(+)E_{N}\,
      E^\p_{N-1}}\right)\rho_{kn}^{I,E_{N}} +
  \left(\Gamma_{l,kjjn}^{(-)E_{N}\, E^\p_{N+1}}-\Gamma_{l,kjjn}^{(-)E_{N}\,
      E^\p_{N-1}}\right)\rho_{nk}^{I,E_{N}}.
\end{equation}

\section{Linear and nonlinear transport}
\label{sec:results}

After having obtained the energy spectrum and the eigenstates of the
DWCNT QD system, we can calculate the transition rates,
Eqs.~\eqref{eq:trans-rate-final-n+1}
and~\eqref{eq:trans-rate-final-n-1} and use the Bloch-Redfield
equations for the reduced density matrix to calculate the
transport properties of the system. Here we
present the calculated
results of both linear and nonlinear conductances.

\subsection{Linear conductance}
\label{sec:linear-conductance}

\begin{figure}
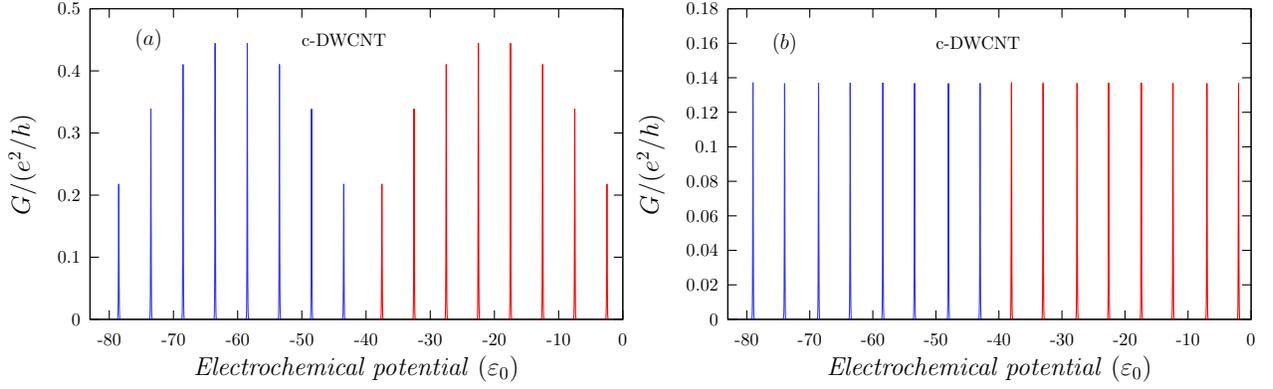

  \includegraphics[width=0.46\textwidth]{fig5a.eps}
  \includegraphics[width=0.46\textwidth]{fig5b.eps}
  \caption{(Color online) Calculated linear conductances as a function
    of the gate 
    electrochemical potential in commensurate double-walled carbon
    nanotube (c-DWCNT) quantum dot (QD) systems with different
    parameters. 
    (a): $\Delta = \zeta = 0.0$, 
    $W_{00}^{++}=W_{00}^{--}=W^{+ -}_{00}/2 = 5.0\varepsilon_0$, and $k_BT = 0.025\varepsilon_0$, where the level
    spacing $\varepsilon_0$ is used as the unit of energy. The coupling 
    strengths are $\gamma_{s\pm}=\gamma_{d\pm}=0.01\varepsilon_0$. (b):
    $\Delta = 0.2$ and $\zeta = 0.3$. 
    The remaining parameters are the same as those in (a).
    In both cases, the linear conductances in c-DWCNTs show an
    8-electron periodicity. In the case of zero mismatch parameters
    shown in (a), an 8-electron periodicity of the heights of the
    conductance peaks also occurs. For finite mismatch shown in (b), the
    peak heights are equal but the 8-electron periodicity of the
    addition energies, i.e., the peak distances, remains, (to
    emphasize this we assign to each ground of eight peaks different
    colors).
  }
  \label{fig:linear-c-com}
\end{figure}

\begin{figure}
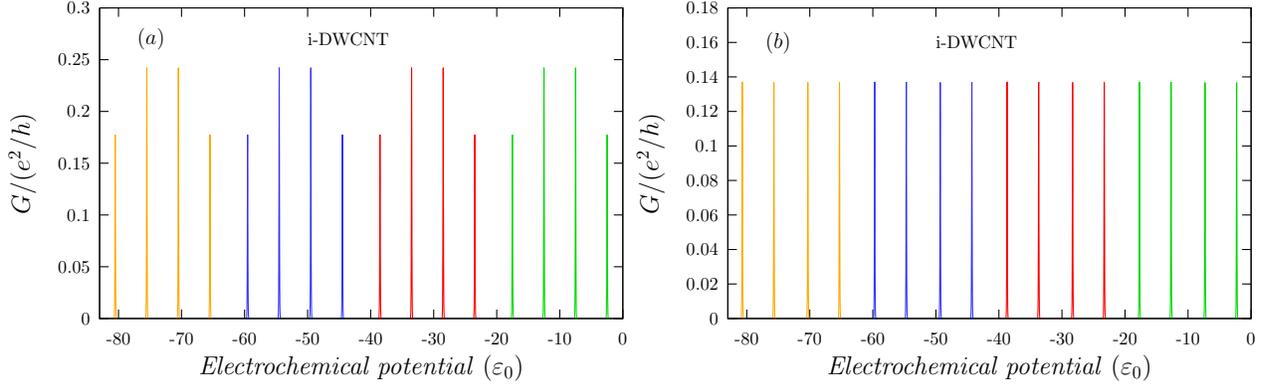

  \includegraphics[width=0.46\textwidth]{fig6a.eps}
  \includegraphics[width=0.46\textwidth]{fig6b.eps}
  \caption{(Color online) Calculated linear conductances as a function
    of the gate 
    electrochemical potential in incommensurate double-walled carbon
    nanotube (i-DWCNT) quantum dot (QD) systems with different
    parameters.
    (a) $\Delta_+ = \Delta_-=0.0$, 
    $W^{++}_{00} = 5\varepsilon_0$, $W^{--}_{00} = 6.0\varepsilon_0$,
    $W^{+ -}_{00} = 5.5\varepsilon_0$, and $k_BT = 0.025\varepsilon_0$, where the level
    spacing $\varepsilon_0$ is used as the unit of energy.
    The coupling strengths are
    $\gamma_{s+}=\gamma_{d+}=0.01\varepsilon_0$ and $\gamma_{s-}=\gamma_{d-}=0$. 
    (b): $\Delta_+ =0.2$ and $\Delta_-=0.3$.
    The remaining parameters are the same as those in (a).  The linear
    conductances in i-DWCNTs show a 4-electron periodicity. In the
    absence of mismatch shown in (a), the 4-electron periodicity is
    observed also in the peak heights while it is no longer observed
    at finite mismatch shown in (b). However, the addition energy,
    i.e., the peak distance, shows a 4-electron periodicity in both
    cases, (to
    emphasize this we assign to each ground of eight peaks different
    colors).
  }
  \label{fig:linear-c-incom}
\end{figure}

In the linear transport regime, i.e., $\abso{eV_b} \ll k_BT \ll
\varepsilon_0$, where $V_b$ is
the applied bias, only the ground states
with $N$ and $N+1$ electrons
are involved in the transport. In this case, the equations for
the diagonal elements and the off-diagonal elements of the reduced
density matrix are decoupled from
each other and we only have to take into account the diagonal
elements of the ground states with a certain electron number, which
are the occupation probabilities. The stationary occupation
probability of the 
ground state with $N$ electrons is given
as~\cite{mayrhofer:epjb2007}
\begin{equation}
  \label{eq:P-solution}
  P_N = \frac{\sum_{l\alpha} \gamma_{l\alpha} (1-f(\varepsilon_l)) C_{N+1,N}^{\alpha} }{\sum_{l\alpha} \gamma_{l\alpha} f(\varepsilon_l) C_{N, N+1}^{\alpha}  +
    \gamma_{l\alpha} ( 1- f(\varepsilon_l)) C_{N+1,N}^{\alpha}},
\end{equation}
where  $C_{N, N+1}^{\alpha} $ are the number of permitted ground states with
$N+1$ particles when one electron is added to a ground state with $N$
particles and this electron is added to the bonding/anti-bonding state
$\alpha$ in c-DWCNTs or 
to the shell $\alpha$ in i-DWCNTs. We define the energy $\varepsilon_l = eV_l - \Delta E$
and the energy difference $\Delta E =  E_N^0 - E_{N+1}^0 - \mu_g$, where $\mu_g$ is the electrochemical
potential in the 
gate. 
The linear conductance is then given as
\begin{equation}
  \label{eq:current-low-e}
  \begin{split}
    G  = \frac{2e^2\beta}{h}
   \frac{\sum_{\alpha} \gamma_{s\alpha}\gamma_{d\alpha} C_{N,N+1}^{\alpha}C_{N+1,N}^{\alpha}}{\sum_{l\alpha}
     \gamma_{l\alpha} f(-\Delta E) C_{N, N+1}^{\alpha} +
    \gamma_{l\alpha} ( 1- f(-\Delta E)) C_{N+1,N}^{\alpha}} \frac{e^{-\beta\Delta E}}{(1 + e^{-\beta\Delta E})^2}, \\
  \end{split}
\end{equation}
where we assume that the bias is symmetrically applied to the source
and drain leads, that is, $-V_s = V_d = V_b/2$. The maximum
value of the linear conductance is
\begin{equation}
  \label{eq:conductance-max}
  G_{\max} = \frac{2e^2\beta}{h} \frac{\sum_{\alpha} \gamma_{s\alpha}\gamma_{d\alpha}
    C_{N,N+1}^{\alpha} C_{N+1,N}^{\alpha}}{\sum_{l\alpha} \gamma_{l\alpha} (C_{N,N+1}^{\alpha} +
    C_{N+1,N}^{\alpha}) +
    2\sqrt{\bigl(\sum_{l\alpha} \gamma_{l\alpha} C_{N,N+1}^{\alpha}
      \bigr) \bigl( \sum_{l\alpha}
      \gamma_{l\alpha} C_{N+1,N}^{\alpha} \bigr)}}.
\end{equation}
and the maxima of the conductance as a function of $\mu_g$ are
at $-\mu_g = E_{N+1}^0 - E_N^0 + \Delta E_{\max}$, where~\cite{glazman:jetpl1988, beenakker:prb1991, alhassid:rmp2000,
  mayrhofer:prb2006, mayrhofer:epjb2007}
\begin{equation}
  \label{eq:peak}
  \Delta E_{\max} = \frac{1}{2\beta} \ln \frac{\sum_{l\alpha} \gamma_{l\alpha} C_{N+1,N}^{\alpha} }{
    \sum_{l\alpha} \gamma_{l\alpha} C_{N,N+1}^{\alpha}}.
\end{equation}
The conductance peak
occurs whenever an electron is added or removed from the DWCNT QD by
changing the electrochemical potential in the gate.
 At zero temperature, from Eq.~\eqref{eq:peak} $\Delta E_{\max}$ vanishes
 and the conductance peak
 occurs when the electrochemical potential of the gate satisfies
 the following condition,
\begin{equation*}
   -\mu_g = E_{N+1}^0 - E_{N}^0 \equiv  \mu_N.
\end{equation*}
Therefore, at zero temperature the addition energy $\delta \mu_N$ is given by
\begin{equation*}
  \delta \mu_N = \abso{\mu_N - \mu_{N-1}} = \abso{E_{N+1}^0 + E_{N-1}^0 - 2E_{N}^0}.
\end{equation*}
For a c-DWCNT QD system, electrons can tunnel into both shells because
of nonzero intershell couplings. Hence there is an 8-electron
periodicity of the conductance peak distances, which are 
  \begin{align}
  \label{eq:distance-cdwcnt}
  \delta \mu_1 &= \delta \mu_3 = \delta \mu_5 = \delta \mu_7 = W^{++}_{00}, \\
  \delta \mu_2 &= \mu_6 = 2 \min(\Delta,\zeta) \varepsilon_0 + W^{++}_{00},  \\
  \delta \mu_4 &= 2 \abso{\Delta-\zeta} \varepsilon_0 + W^{++}_{00}, \\
  \delta \mu_8 &= \varepsilon_0 - 2(\Delta+\zeta)\varepsilon_0 +W^{++}_{00} .
  \end{align}
Here, we use the relation,
$W_{00}^{++}=W_{00}^{--}=W_{00}^{+ -}/2$, in c-DWCNTs
(cf. Eq.~\eqref{eq:U-com}).
On the other hand, electrons can only tunnel into the outer shell
in an i-DWCNT QD system because the contacts are deposited onto the
outer shell and the intershell couplings vanish. Therefore, there is a
4-electron periodicity 
of the conductance peak distance like in a SWCNT QD system, which are
  \begin{align}
  \label{eq:distance-idwcnt}
  \delta \mu_1 &= \delta \mu_3 =W^{++}_{00}, \\
  \delta \mu_2 &=  2 \Delta_+\varepsilon_0 + W^{++}_{00},  \\
  \delta \mu_4 &= \varepsilon_0 - 2\Delta_+\varepsilon_0 + W^{++}_{00}.
  \end{align}
Because electrons tunnel only into the outer shell with the
interaction strength $W_{00}^{++}$, the addition energy $\delta\mu_N$ does not
depend either on the 
interaction strength in the inner shell $W_{00}^{--}$ nor on the
intrashell interaction strength $W_{00}^{+ -}$.
The calculated linear conductances of DWCNT QDs of different
configurations are shown in
Figs.~\ref{fig:linear-c-com} and~\ref{fig:linear-c-incom}. In a
c-DWCNT QD, the intraband interaction 
strengths $W_{00}^{++}$ and $W_{00}^{--}$ are the same while the interband
interaction strength $W_{00}^{+ -}$ is twice as large
(cf. Eq.~\eqref{eq:U-com}) . However, in an i-DWCNT the interaction
strength in the inner shell 
$W_{00}^{--}$
is the strongest because of the smaller inner shell radius and one has
$W_{00}^{--} > W_{00}^{+ -} > W_{00}^{++}$. 
The shapes of the conductance peaks strongly
depend on the mismatch parameters, i.e., $\Delta$ and $\zeta$ in c-DWCNTs and
$\Delta_{\pm}$ in i-DWCNTs. For zero mismatch
parameters, the quantities $C_{N,N+1}^{\alpha}$ and $C_{N+1,N}^{\alpha}$ are
$C_{N,N+1}^{\alpha} = 4,3,2,1$ and $C_{N+1,N}^{\alpha} = 1, 2, 3, 4$ for
$N^\alpha=4m, 4m+1, 4m+2, 4m+3$ with an integer $m$, 
where $N^{\alpha}$ is the electron number either in the bonding/anti-bonding state
$\alpha$ in c-DWCNTs or in the shell $\alpha$ in i-DWCNTs.
Therefore, according to Eq.~\eqref{eq:conductance-max} one can find
that the conductance peak heights show an 8-electron periodicity
in a c-DWCNT QD (cf. Fig.~\ref{fig:linear-c-com}(a)) because both bonding
and anti-bonding states contribute to the electron transport. However,
there is a 4-electron
periodicity in an i-DWCNT QD (cf. Fig.~\ref{fig:linear-c-incom}(a)) because
only the outer shell contributes. 
If
the mismatch parameters are nonzero, we find 
$C_{N,N+1}^\alpha =
2,1,2,1$ and $C_{N+1,N}^{\alpha}=1,2,1,2$ for
$N^\alpha=4m,4m+1,4m+2,4m+3$.
Therefore, all the
conductance peaks have the same heights
(cf. Fig.~\ref{fig:linear-c-com}(b)
and~\ref{fig:linear-c-incom}(b)). However, the distance between two
conductance peaks, i.e., the addition energy, always shows
an 8-electron periodicity in c-DWCNT QDs and
the 4-electron periodicity in i-DWCNTs 
as shown in the Figs.~\ref{fig:linear-c-com}
and~\ref{fig:linear-c-incom}.



\subsection{Nonlinear conductances}
\label{sec:nonlinear-conductances}

\begin{figure}
  \includegraphics[width=0.6\textwidth]{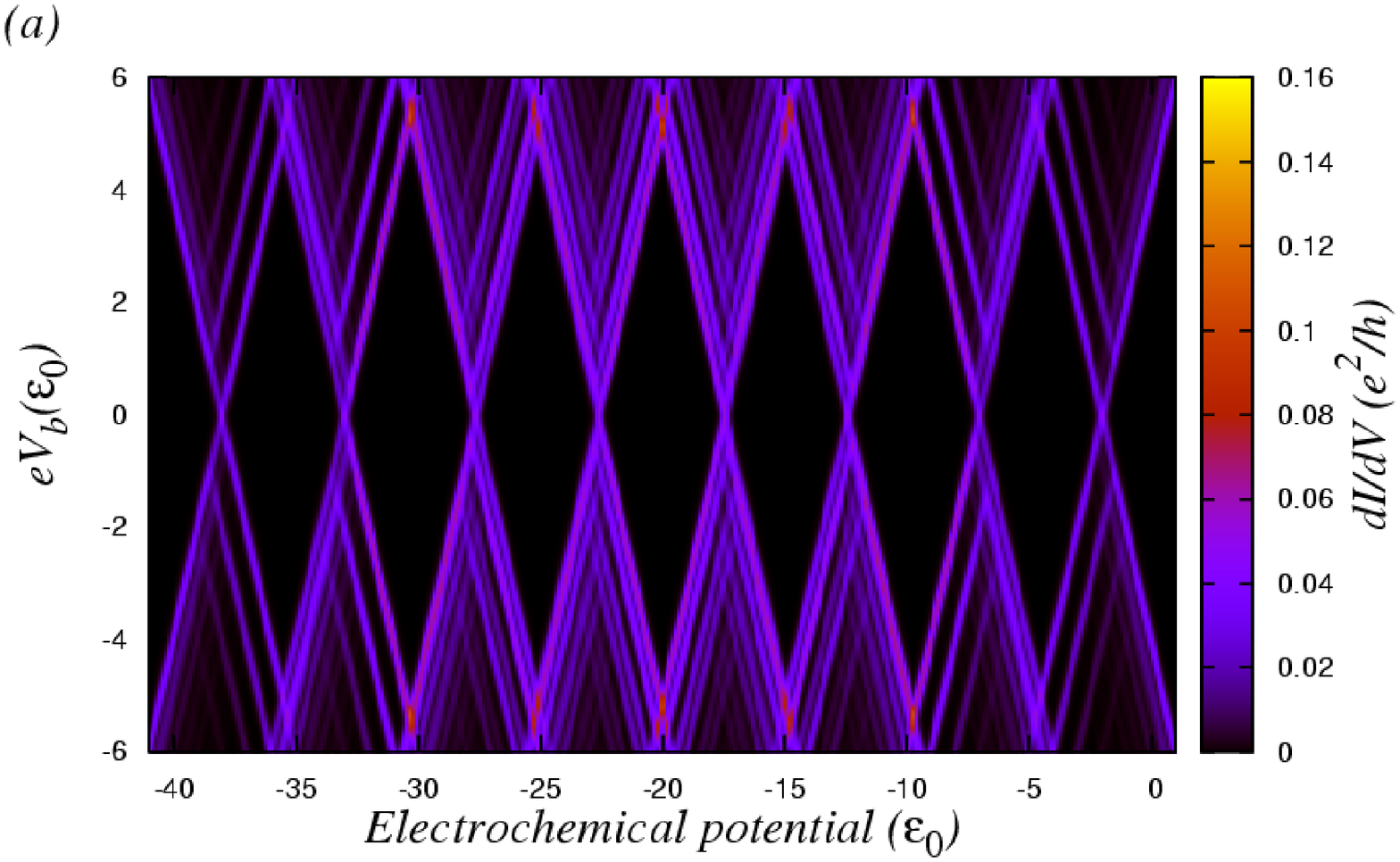}
  \includegraphics[width=0.6\textwidth]{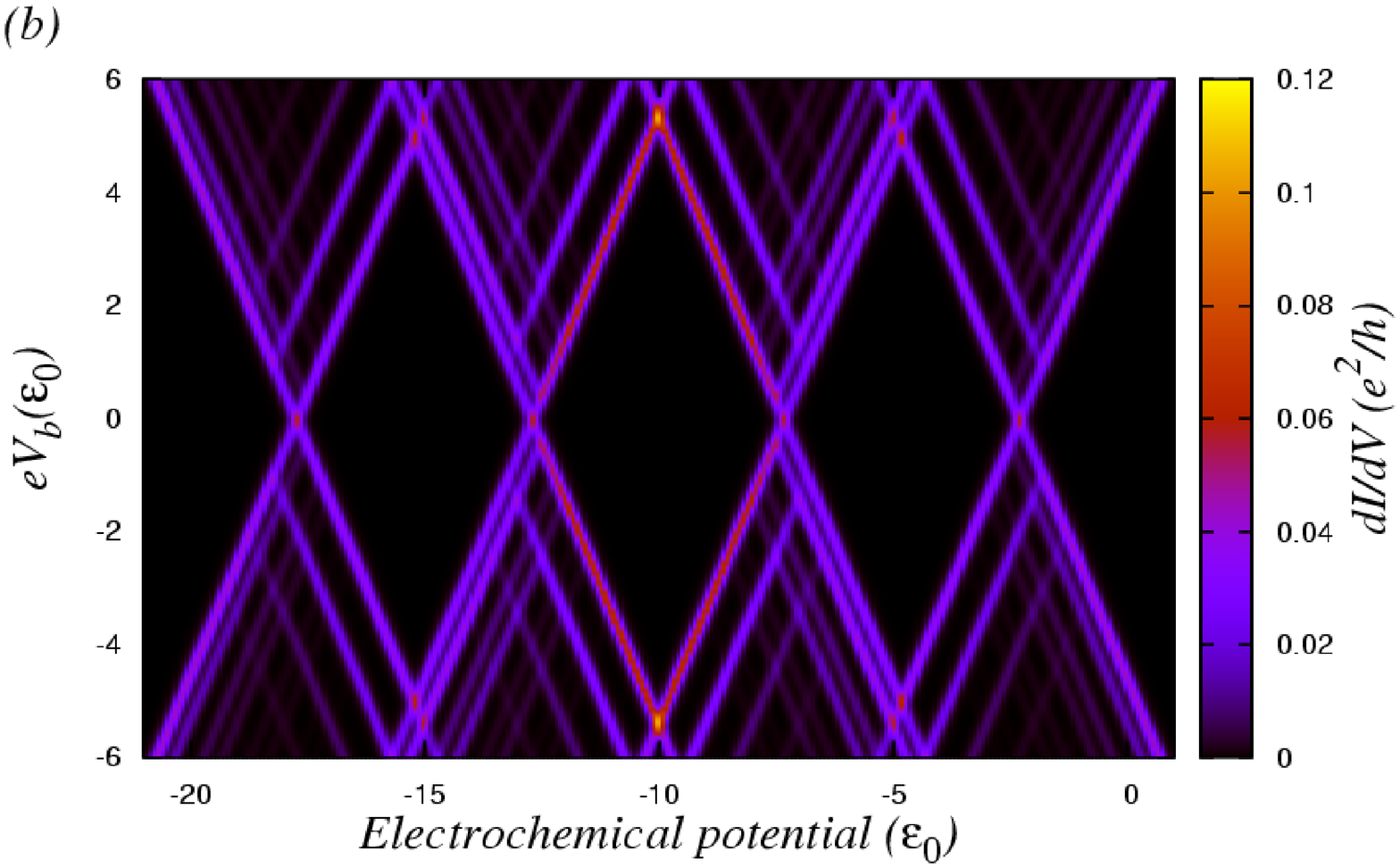}
  \caption{(Color online) Calculated stability diagrams of
    double-walled carbon 
    nanotube (DWCNT) quantum dot (QD) systems. (a) Stability diagram of
    a commensurate DWCNT(c-DWCNT) QD. The parameters are 
    $\Delta = 0.2$, $\zeta = 0.3$, $W^{++}_{00} = W^{--}_{00} =
    W^{+ -}_{00}/2 = 5.0\varepsilon_0$ and $k_BT = 0.05\varepsilon_0$, where the level
    spacing $\varepsilon_0$ is used as the unit of
    energy. We use $\gamma_{s\pm}=\gamma_{d\pm}=0.01\varepsilon_0$
    for the coupling strengths between the leads and the DWCNT
    QD.
    (b) Stability diagram of an
    incommensurate DWCNT(i-DWCNT) QD. The parameters are 
    $\Delta_+ = 0.2$, $\Delta_-=0.3$, 
    $W^{++}_{00} = 5\varepsilon_0$, $W^{--}_{00} = 6.0\varepsilon_0$,
    $W^{+ -}_{00} = 5.5\varepsilon_0$ and $k_BT = 0.05\varepsilon_0$.  The coupling strengths are
    $\gamma_{s+}=\gamma_{d+}=0.01\varepsilon_0$ and $\gamma_{s-}=\gamma_{d-}=0$. 
    The size of the Coulomb
    diamonds shows an 8-electron periodicity in c-DWCNT QDs as shown in
    (a) while it shows a 4-electron periodicity in i-DWCNT QDs as shown
    in (b). 
  }
  \label{fig:stability}
\end{figure}

When higher bias is applied, i.e., $\abso{eV_b} \geq \varepsilon_0 \gg k_BT$, we can
only solve the 
Bloch-Redfield equations numerically. 
For elastic tunneling process, we have to
include the coherences between the states with same particle number
$\ve{N}$ but with different bosonic excitations
$\ve{m}$~\cite{mayrhofer:prb2006, mayrhofer:epjb2007}. Because of
the large number of degenerate bosonic excitations, the rank
of the reduced density matrix increases very fast as the applied bias
increases, which causes a very long computing time to solve the 
equations. 
On the other hand, these
coherences can be ignored in an inelastic tunneling process, in which the QD
system will be restored to the equilibrium states before the next tunneling
process. Only the diagonal elements in the reduced density matrix are
nonzero and they obey the Boltzmann distribution as
\begin{equation*}
  \rho_{nn}^{I,E_{N}} (t) = \mP_N(t) \frac{e^{-\beta E_{N}^n}}{\sum_{k} e^{-\beta E_{N}^k}},
\end{equation*}
where $n$ and $k$ are indices of the eigenstates of the DWCNT QD
Hamiltonian and $\mP_N(t)$ is the probability of finding $N$
electrons in the QD. Instead of solving the Bloch-Redfield equations
directly,
we can solve the equation of motion for the probability $\mP_N(t)$,
\begin{equation}
  \label{eq:eom-P-inelastic}
  \frac{d}{dt} \mP_N(t) = - \sum_{l, M=N\pm 1} \Theta_l^{N\to M}  + \sum_{l,M=N\pm 1} \Theta_l^{M\to N},
\end{equation}
where the tunneling rate is defined in Eq.~\eqref{eq:tunnel-rate} and
can now be expressed in terms of $\mP_N(t)$ as
\begin{equation}
  \label{eq:tunnel-rate-inelastic}
  \Theta_l^{N\to N\pm 1} = \mP_N(t)  \sum_{E,E^\p} \frac{e^{-\beta E_{N}^n}}{\sum_{k} e^{-\beta E_{N}^k}} 
  \Bigl( \sum_{nj} \Gamma_{l,njjn}^{(+)E_N \,E_{N\pm 1}^\p} + \Gamma_{l,njjn}^{(-)E_N\,
    E_{N\pm1}^\p} \Bigr).
\end{equation}
The number of the equations reduces significantly and the equations
can be solved quite fast.
In Fig.~\ref{fig:stability}, we show the calculated stability
diagram of a DWCNT QD system in an inelastic tunneling process.
The size of the Coulomb diamonds shows an
8-electron periodicity in c-DWCNT QDs and an 4-electron
periodicity in i-DWCNT QDs. The excitation lines are also shown in
Fig.~\ref{fig:stability}, which contain contributions of both
fermionic 
(cf. Eqs.~\eqref{eq:H-groundstate-cdwcnt} and~\eqref{eq:H-groundstate-idwcnt})
and bosonic excitations
(cf. Eq.~\eqref{eq:H-ex-diagonalized}). There are more excitation 
lines in c-DWCNT QDs than in i-DWCNT QDs because of the larger
number of the ground states of c-DWCNTs. The stability diagram
of an i-DWCNT QD looks quite similar to that of a SWCNT QD, which
shows
also a 4-electron periodicity. However, the configuration of the excitation
lines of the two cases are different, because 
the excitation spectrum in i-DWCNTs contains an extra contribution
from the 
Coulomb interaction due to the electrons in the inner shell.


\section{Conclusions}
\label{sec:conclusions}

In this paper, we derive the energy spectrum of both finite length
c-DWCNTs and 
i-DWCNTs with open boundary conditions. Then we
calculate the transport properties of the DWCNT QD system by solving
the Bloch-Redfield equations for the reduced density matrix of the QD
systems. Because the contacts are usually deposited on the outer shell and
the intershell coupling depends on the chiralities of the two shells, we
find an 8-electron periodicity of the linear conductance peak
distances in c-DWCNTs but a 4-electron periodicity in i-DWCNTs. The peak
heights strongly depend on the degeneracies of the ground states.
By including both fermionic and bosonic excitations, we also
calculate the stability diagrams of QD systems with both c-DWCNTs and
i-DWCNTs in an inelastic tunneling process. 
The periodicity
of the Coulomb diamond sizes depends on the number of the shells
contributing to the electron transport.
Therefore, the 4-electron periodicity 
in a MWCNT QD
measured in the experiments in Ref.~\onlinecite{buitellar:prl2002} may
be due to the fact that only the outermost metallic shell was
involved in the electron transport. 
Similarly, the 4-electron periodicity
in i-DWCNTs is because of the negligible intershell coupling and there
will be an 8-electron periodicity if a large intershell
coupling is caused for example by contacts. 
Therefore, it's necessary to use properly prepared contacts in
order to observe the different periodicities of Coulomb blockade oscillations
in different types of DWCNTs in the experiments. 


\begin{acknowledgments}
  The authors would like to thank L. Mayrhofer for helpful discussions. The
  authors acknowledge the support of DFG under the program GRK 638. 
\end{acknowledgments}

\appendix

\section{Matrix elements of the electron operators}
\label{sec:matrix-e-op}

In this appendix, we calculate
the matrix elements of the electron operators
Eqs.~\eqref{eq:e-op-idwcnt} and~\eqref{eq:e-op-cdwcnt} in
the basis of the eigenstates, Eq.~\eqref{eq:Hex-states}, of the
Hamiltonian $H_{QD}$ and the results are
used in Sec.~\ref{sec:dynamics-qd}. The matrix elements of the
electron operator are calculated by using the relations
\begin{equation}
  \label{eq:element-e-3d}
  \begin{split}
    \bigl\langle \ve{N}, \ve{m} \bigl\lvert \Psi_{\alpha \sigma}(\ve{r}) \bigr\rvert
    \ve{N}^\p, \ve{m}^\p \bigr\rangle 
    = \sqrt{L} \sum_{\rt F} \sgn(F) \varphi_{\alpha \sgn(F)\rt F}(\ve{r})
    \bigl\langle \ve{N}, \ve{m} \bigl\lvert \psi_{\alpha \rt\sigma F}(u) \bigr\rvert
    \ve{N}^\p, \ve{m}^\p \bigr\rangle.
  \end{split}
\end{equation}
The 1D electron operator $\psi_{\alpha \rt F \sigma}$ can be expressed in terms of the
bosonic operators introduced above as~\cite{voit:rpp1994, delft:adp1998}
\begin{equation}
  \label{eq:1d-e-op-boson}
  \ann{\psi}{\alpha\rt F\sigma }(u) = \frac{\ann{\eta}{\alpha \rt\sigma} K_{\alpha\rt F\sigma
    }(u)}{\sqrt{1-e^{-a\pi/L}}} e^{i\cre{\phi}{\alpha \rt F \sigma }(u) + i \ann{\phi}{\alpha
      \rt F \sigma }(u)},
\end{equation}
where $a$ is an infinitesimal positive number used to avoid the
divergence in the long wave length limit and the operator
$\ann{\eta}{\alpha\rt\sigma}$ is the Klein factor, which 
destroys a particle in the branch $\alpha\rt \sigma$ when acting on the
eigenstates of the DWCNT Hamiltonian,
\begin{equation*}
  \ann{\eta}{\alpha \rt\sigma} \ket{\ve{N}, \ve{m}} = (-1)^{\sum_{i=1}^{\alpha \rt\sigma -1} N_i}
  \ket{\ve{N} - \ve{e}_{\alpha \rt\sigma}, \ve{m}},
\end{equation*}
where we use the convention $i = + \Rt \uparrow , + \Lt \uparrow , + \Rt \downarrow , + \Lt \downarrow
, - \Rt \uparrow , - \Lt \uparrow , - \Rt \downarrow , - \Lt \downarrow \; = 1, 2, 3, 4, 5, 6, 7, 8$ and
the vector $\ve{e}_{\alpha\rt\sigma}$ denotes a state where there is only one
particle in the branch $\alpha\rt\sigma$. The notation $\sum_{i=1}^{\alpha\rt\sigma-1}$ means
that the sum runs over all the state from $1$ to $i=\alpha\rt\sigma -1$
with $\alpha\rt\sigma$ fixed by the unit vector $\ve{e}_{\alpha\rt\sigma}$.
The phase factor is
\begin{equation*}
  K_{\alpha \rt F \sigma }(u) = \frac{1}{\sqrt{2L}} e^{i\frac{\pi}{L} \sgn(F)
    (\sgn(\rt)N_{\alpha  \rt\sigma} + \Delta_{\alpha})u},
\end{equation*}
where $\Delta_{\pm}=\Delta$ for the c-DWCNTs.
The field operator $\cre{\phi}{\alpha \rt F \sigma}$ is given as
\begin{equation*}
  i\ann{\phi}{\alpha \rt F \sigma}(u) = \sum_{q>0} \frac{e^{-a q/2}}{\sqrt{n_q}}
  e^{i\sgn(\rt F)qu} b_{\alpha \sgn(\rt)q \sigma}.
\end{equation*}
Therefore, the matrix elements of the 1D electron operator have the
form~\cite{mayrhofer:epjb2007}
\begin{equation}
  \label{eq:element-1d-e-op}
  \bigl\langle \ve{N}, \ve{m} \bigl\lvert \psi_{\alpha \rt F \sigma }(u) \bigr\rvert
  \ve{N}^\p, \ve{m}^\p \bigr\rangle = \delta_{\ve{N}+\ve{e}_{\alpha\rt\sigma}, \ve{N}^\p}
  \frac{(-1)^{\sum_{i}^{\alpha\rt\sigma -1} N_i}}{\sqrt{1-e^{-a\pi/L}}} K_{\alpha\rt F \sigma
  }(u) \prod_{q>0}\prod_{j\delta\xi } F(\lambda_{\alpha\rt\sigma q}^{j\delta\xi F}(u), m_{j\delta\xi q}, m_{j\delta\xi
    q}^\p),
\end{equation}
where the function $F$ can be expressed in terms of  
the Laguerre polynomials $L_m^n$~\cite{gradshteyn:2000}
\begin{equation}
  \label{eq:F-function}
  F(\lambda, m, m^\p) = \frac{m_{\mmin}!}{m_{\mmax}!}
  L_{m_{\mmin}}^{m_{\mmax} - m_{\mmin}} (\abso{\lambda}^2) \Bigl( \Theta(m^\p -
  m) \lambda^{m^\p - m} + \Theta(m - m^\p) (-\lambda^\ast)^{m-m^\p} \Bigr)
\end{equation}
with $m_{\mmax} = \max(m, m^\p)$ and $m_{\mmin} = \min(m,
m^\p)$. $\Theta(x)$ is the Heaviside step function and the
parameters $\lambda$'s are given by
\begin{equation}
  \label{eq:lambda-2}
  \lambda_{\alpha \rt\sigma q}^{j\delta\xi F}(u) = \frac{\Lambda_{\alpha \rt \sigma }^{j\delta\xi q}}{\sqrt{n_q}}
  \bigl(e^{i\sgn(\rt F)qu} S_{j\delta \xi q} - e^{-i\sgn(\rt F) qu} C_{j\delta\xi q}
  \bigr).
\end{equation}

\section{Expressions of the tunneling rates}
\label{sec:tunneling-rate}

In this appendix, we give a derivation of the expressions of the
tunneling rates,
Eqs.~\eqref{eq:trans-rate-final-n+1}
and~\eqref{eq:trans-rate-final-n-1}.
In general, the expressions of the tunneling rates $\Gamma$'s are given
by~\cite{mayrhofer:epjb2007},
\begin{align}
  \label{eq:transition-rate-n+1}
  \Gamma_{l\, k^\p mnk}^{(\pm)E_{N}\, E_{N+1}} &= \frac{1}{\hbar^{2}}\sum_{\alpha \sigma} \iint
  d \ve{x} d \ve{y}  \Bigl( \ann{\Psi}{\alpha \sigma} (\ve{x}) \Bigr)_{k^\p
    m}^{E_{N}\, E^\p_{N+1}} \Bigl(\cre{\Psi}{\alpha \sigma}(\ve{y})
  \Bigr)_{nk}^{E^\p_{N+1}\, E_{N}}  \nonumber \\
  &\qquad \times \int_{0}^{\infty} dt^\p \, \mF_{l\alpha  \sigma }
  (\ve{x},\ve{y},t') e^{\mp \frac{i}{\hbar} \left(E^\p_{N+1}-E_{N} \right)
    t^\p}, \\
  \label{eq:transition-rate-n-1}
  \Gamma_{l\, k^\p mnk}^{(\pm)E_{N} \, E_{N-1}} &= \frac{1}{\hbar^{2}}\sum_{\alpha \sigma} \iint d
  \ve{x} d \ve{y} 
  \left(\cre{\Psi}{\alpha \sigma} (\ve{x}) \right)_{k^\p m}^{E_{N}\, E^\p_{N-1}}
  \left(\ann{\Psi}{\alpha \sigma} (\ve{y}) \right)_{nk}^{E^\p_{N-1}\, E_{N}}
  \nonumber \\
  &\qquad \times \int_{0}^{\infty} dt^\p \mE_{l\alpha \sigma}
  (\ve{x},\ve{y},t^\p)e^{\mp \frac{i}{\hbar}\left(E^\p_{N-1}-E_{N}\right)t^\p},
\end{align}
where $t^\p = t - t_1$. The two functions are defined as
\begin{align}
  \label{eq:trans-E}
  \mE_{l\alpha \sigma}(\ve{x},\ve{y},t^\p)&= T_{l \alpha }(\ve{x})T_{l \alpha }^{*}(\ve{y})\left\langle
    \ann{\Phi}{l\sigma }(\ve{x}) \cre{\Phi}{l\sigma}(\ve{y},-t^\p)\right\rangle_{\mathrm{th}}
  \nonumber \\
  &= T_{l\alpha }(\ve{x})T_{l\alpha}^{*}(\ve{y})\int d\varepsilon g_l(\varepsilon)(1-f(\varepsilon)) \sum_{\ve{q}}
  \phi_{\ve{q}}(\ve{x})
  \phi_{\ve{q}}^{*}(\ve{y})e^{-\frac{i}{\hbar}(\varepsilon-eV_{l})t^\p}, \\
  \label{eq:trans-F}
  \mathcal{F}_{l\alpha\sigma}(\ve{x},\ve{y},t^\p) &=
  T_{l\alpha}^{*}(\ve{x})T_{l\alpha}(\
  ve{y})
  \left\langle \cre{\Phi}{l\sigma}(\ve{x}) \ann{\Phi}{l\sigma}(\ve{y},-t^\p)\right\rangle_{\mathrm{th}}
  \nonumber \\
  &= T_{l\alpha}^{*}(\ve{x})T_{l\alpha}(\ve{y})\int
  d\varepsilon
  g_l(\varepsilon)f(\varepsilon)\sum_{\ve{q}}\phi_{\ve{q}}^{*}(\ve{x})\phi_{\ve{q}}(\ve{y})e^{\frac{i}{\hbar}(\varepsilon-eV_{l})t^\p
  },
\end{align}
where $g_{l}(\varepsilon)$ is the density of states in lead $l$, $V_l$ is the
voltage in the lead $l$, and $f(\varepsilon)$ is
the Fermi distribution function. 
In the
following we shall derive the expressions for the tunneling rates
$\Gamma_{l\, k^\p mnk}^{(\pm)E_N \, E_{N+1}}$ for an i-DWCNT QD system as an
example and the 
expressions for the other tunneling rates 
can be obtained by the same
method. By substituting Eqs.~\eqref{eq:trans-F}
and~\eqref{eq:e-op-idwcnt} 
into Eq.~\eqref{eq:transition-rate-n+1}, we have
\begin{equation}
  \label{eq:tunneling-rate-1}
  \begin{split}
    \Gamma_{l\, k^\p mnk}^{(\pm)E_N \, E_{N+1}} &= \frac{\pi L}{\hbar} \sum_{\alpha \sigma}
    \sum_{\rt\rt^\p} \sum_{FF^\p} \sgn(FF^\p) 
    \iint d \ve{x} d \ve{y} \;
    \varphi_{\alpha\sgn(F)\rt F}(\ve{x}) \varphi_{\alpha\sgn(F^\p)\rt^\p F^\p}^\ast(\ve{y}) \\
    &\qquad \times g_l(\varepsilon_l)f(\varepsilon_l) T_{l\alpha}^\ast(\ve{x}) T_{l\alpha}(\ve{y}) \sum_{\ve{q}}
    \phi_{\ve{q}}^\ast(\ve{x}) \phi_{\ve{q}}(\ve{y}) \\
    &\qquad \times \Bigl( \ann{\psi}{\alpha\rt\sigma F} (u_l) \Bigr)_{k^\p
      m}^{E_{N}\, E^\p_{N+1}} \Bigl(\cre{\psi}{\alpha\rt^\p\sigma F^\p}(u_l)
    \Bigr)_{nk}^{E^\p_{N+1}\, E_{N}}, 
  \end{split}
\end{equation}
where $\varepsilon_l = eV_l - E_N - E_{N+1}$ and $u_l = 0, L$ for $l=s, d$. We ignore the slow
oscillations of the 1D electron operators 
along the length of the tunneling interfaces and therefore the product
$ \Bigl( \ann{\psi}{\alpha\rt\sigma F} (u_l) \Bigr)_{k^\p
  m}^{E_{N}\, E^\p_{N+1}} \Bigl(\cre{\psi}{\alpha\rt^\p\sigma F^\p}(u_l)
\Bigr)_{nk}^{E^\p_{N+1}\, E_{N}}$ is independent of the positions.
The integrals over $\varepsilon$ and
$t^\p$ are carried out by using
\begin{equation*}
  \int d\varepsilon g_l(\varepsilon) \int_0^{\infty} dt^\p e^{\pm\frac{i}{\hbar}(\varepsilon-E)t^\p} = \pi\hbar g(E) \pm i\hbar
  \mathcal{P} \int \frac{g(\varepsilon)}{\varepsilon - E} d\varepsilon
\end{equation*}
with $\mathcal{P}$ denotes the Cauchy principal value. We assume that
the width of the lead energy band is infinite and that
the lead density of states $g_l(\varepsilon)$ is constant. 
Hence the
Cauchy principal value is zero. Let's focus on the part depending on
the position in Eq.~\eqref{eq:tunneling-rate-1}, namely,
\begin{equation}
  \label{eq:rate-x}
  I = \iint d \ve{x} d \ve{y} \;
  \varphi_{\alpha\sgn(F)\rt F}(\ve{x}) \varphi_{\alpha\sgn(F^\p)\rt^\p F^\p}^\ast(\ve{y})
  T_{l\alpha}^\ast(\ve{x}) T_{l\alpha}(\ve{y}) \sum_{\ve{q}}
  \phi_{\ve{q}}^\ast(\ve{x}) \phi_{\ve{q}}(\ve{y}). 
\end{equation}
Because the Bloch waves
$\varphi_{\alpha\sgn(F)\rt F}$ from Eq.~\eqref{eq:periodic-f} are largely
localized around the carbon atoms and on the length scale of the $p_z$
orbitals all the other quantities in Eq.~\eqref{eq:rate-x}
are slowly varying, we can rewrite the two integrals as two sums over the
positions of the carbon atoms and the Eq.~\eqref{eq:rate-x} becomes
\begin{equation}
  \label{eq:rate-x2}
  I = \frac{C}{N_\alpha} \sum_{\ve{R}p} \sum_{\ve{R}^\p p^\p}  
  e^{i \ve{F} \cdot \ve{R}} e^{i \ve{F}^\p \cdot \ve{R}^\p} f_{\alpha p \sgn(F)\rt F}
  f_{\alpha p\sgn(F^\p)\rt^\p F^\p}^\ast 
  T_{l\alpha}^\ast(\ve{x}_{\ve{R},p}) T_{l\alpha}(\ve{y}_{\ve{R}^\p, p^\p}) \sum_{\ve{q}}
  \phi_{\ve{q}}^\ast(\ve{x}_{\ve{R},p} ) \phi_{\ve{q}}(\ve{y}_{\ve{R}^\p,p^\p}
  ), 
\end{equation}
where the constant $C$ denotes the integration over the $p_z$
orbitals. Because the leads are described by 3D Fermi gases, the wave
functions $\phi_{l\ve{q}}(\ve{x})$ are the plane waves,
\begin{equation*}
  \phi_{l \ve{q}}(\ve{x}) = \frac{1}{\sqrt{V_l}} e^{i\ve{q}\cdot \ve{x}}
\end{equation*}
with the volume of the gas $V_l$. The sum over $\ve{q}$ can be
performed 
\begin{equation*}
  \sum_{\ve{q}} \phi_{\ve{q}}^\ast(\ve{x}) \phi_{\ve{q}}(\ve{y}) \approx \frac{4\pi
    \sin(\abso{\ve{q}} \abso{\ve{x} - \ve{y}})}
  { \abso{\ve{q}} \abso{\ve{x} - \ve{y}}},
\end{equation*}
which is peaked around $\ve{x} = \ve{y}$. Because of the large Fermi
energy in the leads, the above expression can be approximated by two
Kronecker $\delta$'s,
\begin{equation*}
  \sum_{\ve{q}} \phi_{\ve{q}}^\ast(\ve{x}_{\ve{R},p} )
  \phi_{\ve{q}}(\ve{y}_{\ve{R}^\p, p^\p} ) \approx 4\pi \delta_{\ve{R} \ve{R}^\p} \delta_{pp^\p}.
\end{equation*}
Therefore,  Eq.~\eqref{eq:rate-x2} becomes
\begin{equation}
  \label{eq:rate-x3}
  I = \frac{C}{N_\alpha} \sum_{\ve{R}p} 
  e^{(i \ve{F} - \ve{F}^\p) \cdot \ve{R} } f_{\alpha p \sgn(F)\rt F}
  f_{\alpha p\sgn(F^\p)\rt^\p F^\p}^\ast 
  T_{l\alpha}^\ast(\ve{x}_{\ve{R},p}) T_{l\alpha}(\ve{y}_{\ve{R}, p}) .
\end{equation}
Because of the fast oscillating phase $e^{(i \ve{F} - \ve{F}^\p) \cdot
  \ve{R} }$, the quantity $I$ is nonzero only if $F=F^\p$ and in
turn the sum over $p$ can be carried out as
\begin{equation*}
  \sum_{p}  f_{\alpha p \sgn(F)\rt F} f_{\alpha p\sgn(F)\rt^\p F}^\ast  =
  \delta_{\rt\rt^\p},
\end{equation*}
which can be easily verified by using the explicit expressions
Eqs~\eqref{eq:coeff-1} and~\eqref{eq:coeff-2}. Eq.~\eqref{eq:rate-x3}
becomes
\begin{equation}
  \label{eq:rate-x4}
  I = \frac{C}{N_\alpha} \sum_{\ve{R}p} 
  T_{l\alpha}^\ast(\ve{x}_{\ve{R},p}) T_{l\alpha}(\ve{y}_{\ve{R}, p})  \delta_{FF^\p}
  \delta_{\rt\rt^\p}.
\end{equation}
By substituting Eq.~\eqref{eq:rate-x4} and Eq.~\eqref{eq:element-1d-e-op}
of the matrix elements of the electron operators into
Eq.~\eqref{eq:tunneling-rate-1}, we finally obtain the expression of
the tunneling rates,
\begin{equation}
  \label{eq:rate-final-n+1}
  \begin{split}
    \Gamma_{l\, k^\p mnk}^{(\pm)E_{N}\, E_{N+1}} 
    &= \sum_{\alpha} 
    \frac{\gamma_{l\alpha}}{h} g_l(\varepsilon_l) f(\varepsilon_l) \sum_{\rt \sigma F}
    \delta_{\ve{N} + \ve{e}_{\alpha\rt\sigma}, \, \ve{N+1}}
    \prod_{q >0} \prod_{q^\p>0} \prod_{j\delta\xi} \prod_{j^\p\delta^\p\xi^\p}  \\
    &\qquad \times    F(\lambda_{\alpha\rt\sigma q}^{j\delta\xi F} (u_l), k_{j\delta\xi q}^{\p}, m_{j\delta\xi q})
    F^\ast(\lambda_{\alpha\rt\sigma q}^{j^\p\delta^\p\xi^\p F} (u_l), n_{j^\p\delta^\p\xi^\p
      q^\p}, k_{j^\p\delta^\p\xi^\p q^\p}), 
  \end{split}
\end{equation}
where 
the vector $\ve{e}_{\alpha\rt\sigma}$
denotes a state with one particle in the branch $\alpha\rt\sigma$.
The function $F(\lambda, m, m^\p)$ is given in Eq.~\eqref{eq:F-function} and
the parameters $\lambda_{\alpha\rt\sigma}^{j\delta\xi F}$'s are defined in
Eq.~\eqref{eq:lambda-2}. 
The constants $\gamma_{l\alpha}$ describes the coupling strengths
between the shell $\alpha$ in i-DWCNTs and the leads $l$ having the form 
\begin{equation*}
  \gamma_{l\alpha} = \pi^2 C \sum_{\ve{R}p}
  \frac{\abso{T_{l\alpha}(\ve{x}_{\ve{R},p})}^2}{1-e^{-a\pi/L}}, 
\end{equation*}
The four eigenstates are
\begin{align*}
  \ket{k^\p} &= \ket{\ve{N}, \ve{k}^\p},& \ket{m} &= \ket{\ve{N+1},
    \ve{m}}, \\
  \ket{n} &= \ket{\ve{N+1}, \ve{n}}, & \ket{k} &= \ket{\ve{N},
    \ve{k}}. 
\end{align*}
Similarly, the expressions for the other tunneling rates are
\begin{equation}
  \label{eq:rate-final-n-1}
  \begin{split}
    \Gamma_{l\, k^\p mnk}^{(\pm)E_{N}\, E_{N-1}} 
    &= \sum_{\alpha} 
    \frac{\gamma_{l\alpha}}{h} g_l(\varepsilon_l) (1-f(\varepsilon_l)) \sum_{\rt \sigma F}
    \delta_{\ve{N} - \ve{e}_{\alpha\rt\sigma}, \, \ve{N-1}}
    \prod_{q >0} \prod_{q^\p>0} \prod_{j\delta\xi} \prod_{j^\p\delta^\p\xi^\p} \\
    &\qquad \times    F^\ast(\lambda_{\alpha\rt\sigma q}^{j\delta\xi F} (u_l), k_{j\delta\xi q}^{\p}, m_{j\delta\xi q})
    F(\lambda_{\alpha\rt\sigma q}^{j^\p\delta^\p\xi^\p F} (u_l), n_{j^\p\delta^\p\xi^\p
      q^\p}, k_{j^\p\delta^\p\xi^\p q^\p}) 
  \end{split}
\end{equation}
with the eigenstates
\begin{align*}
  \ket{k^\p} &= \ket{\ve{N}, \ve{k}^{\p}},& \ket{m} &= \ket{\ve{N-1},
    \ve{m}}, \\
  \ket{n} &= \ket{\ve{N-1}, \ve{n}}, & \ket{k} &= \ket{\ve{N},
    \ve{k}}. 
\end{align*}
From the expressions of the tunneling rate,
Eqs.~\eqref{eq:rate-final-n+1}
and~\eqref{eq:rate-final-n-1}, we can see clearly that
the contacts do not mix the electrons in different branches.



\begin{thebibliography}{40}
\expandafter\ifx\csname natexlab\endcsname\relax\def\natexlab#1{#1}\fi
\expandafter\ifx\csname bibnamefont\endcsname\relax
  \def\bibnamefont#1{#1}\fi
\expandafter\ifx\csname bibfnamefont\endcsname\relax
  \def\bibfnamefont#1{#1}\fi
\expandafter\ifx\csname citenamefont\endcsname\relax
  \def\citenamefont#1{#1}\fi
\expandafter\ifx\csname url\endcsname\relax
  \def\url#1{\texttt{#1}}\fi
\expandafter\ifx\csname urlprefix\endcsname\relax\def\urlprefix{URL }\fi
\providecommand{\bibinfo}[2]{#2}
\providecommand{\eprint}[2][]{\url{#2}}

\bibitem[{\citenamefont{Ijima}(1991)}]{ijima:nature1991}
\bibinfo{author}{\bibfnamefont{S.}~\bibnamefont{Ijima}},
  \bibinfo{journal}{Nature} \textbf{\bibinfo{volume}{354}}, \bibinfo{pages}{56}
  (\bibinfo{year}{1991}).

\bibitem[{\citenamefont{Saito et~al.}(1998)\citenamefont{Saito, Dresselhaus,
  and Dresselhaus}}]{saito:1998}
\bibinfo{author}{\bibfnamefont{R.}~\bibnamefont{Saito}},
  \bibinfo{author}{\bibfnamefont{G.}~\bibnamefont{Dresselhaus}},
  \bibnamefont{and} \bibinfo{author}{\bibfnamefont{M.~S.}
  \bibnamefont{Dresselhaus}}, \emph{\bibinfo{title}{Physical Properties of
  Carbon Nanotubes}} (\bibinfo{publisher}{Imperial College Press, London},
  \bibinfo{year}{1998}).

\bibitem[{\citenamefont{Charlier et~al.}(2007)\citenamefont{Charlier, Blase,
  and Roche}}]{charlier:677}
\bibinfo{author}{\bibfnamefont{J.-C.} \bibnamefont{Charlier}},
  \bibinfo{author}{\bibfnamefont{X.}~\bibnamefont{Blase}}, \bibnamefont{and}
  \bibinfo{author}{\bibfnamefont{S.}~\bibnamefont{Roche}},
  \bibinfo{journal}{Rev. Mod. Phys.} \textbf{\bibinfo{volume}{79}},
  \bibinfo{eid}{677} (\bibinfo{year}{2007}).

\bibitem[{\citenamefont{Loiseau et~al.}(2006)\citenamefont{Loiseau, Launois,
  Petit, Roche, and Salvetat}}]{loiseau-nt:2006}
\bibinfo{editor}{\bibfnamefont{A.}~\bibnamefont{Loiseau}},
  \bibinfo{editor}{\bibfnamefont{P.}~\bibnamefont{Launois}},
  \bibinfo{editor}{\bibfnamefont{P.}~\bibnamefont{Petit}},
  \bibinfo{editor}{\bibfnamefont{S.}~\bibnamefont{Roche}}, \bibnamefont{and}
  \bibinfo{editor}{\bibfnamefont{J.}~\bibnamefont{Salvetat}}, eds.,
  \emph{\bibinfo{title}{Understanding Carbon Nanotubes}}, vol.
  \bibinfo{volume}{677} of \emph{\bibinfo{series}{Lecture Notes in Physics}}
  (\bibinfo{publisher}{Springer Berlin}, \bibinfo{year}{2006}).

\bibitem[{\citenamefont{Egger and Gogolin}(1997)}]{egger:prl1997}
\bibinfo{author}{\bibfnamefont{R.}~\bibnamefont{Egger}} \bibnamefont{and}
  \bibinfo{author}{\bibfnamefont{A.~O.} \bibnamefont{Gogolin}},
  \bibinfo{journal}{Phys. Rev. Lett.} \textbf{\bibinfo{volume}{79}},
  \bibinfo{pages}{5082} (\bibinfo{year}{1997}).

\bibitem[{\citenamefont{Egger and Gogolin}(1998)}]{egger:epjb1998}
\bibinfo{author}{\bibfnamefont{R.}~\bibnamefont{Egger}} \bibnamefont{and}
  \bibinfo{author}{\bibfnamefont{A.~O.} \bibnamefont{Gogolin}},
  \bibinfo{journal}{Eur. Phys. J. B} \textbf{\bibinfo{volume}{3}},
  \bibinfo{pages}{281} (\bibinfo{year}{1998}).

\bibitem[{\citenamefont{Kane et~al.}(1997)\citenamefont{Kane, Balents, and
  Fisher}}]{kane:prl1997}
\bibinfo{author}{\bibfnamefont{C.}~\bibnamefont{Kane}},
  \bibinfo{author}{\bibfnamefont{L.}~\bibnamefont{Balents}}, \bibnamefont{and}
  \bibinfo{author}{\bibfnamefont{M.~P.~A.} \bibnamefont{Fisher}},
  \bibinfo{journal}{Phys. Rev. Lett.} \textbf{\bibinfo{volume}{79}},
  \bibinfo{pages}{5086} (\bibinfo{year}{1997}).

\bibitem[{\citenamefont{Bockrath et~al.}(1999)\citenamefont{Bockrath, Cobden,
  Lu, Rinzler, Smalley, Balents, and McEuen}}]{bockrath:nature1999}
\bibinfo{author}{\bibfnamefont{M.}~\bibnamefont{Bockrath}},
  \bibinfo{author}{\bibfnamefont{D.~H.} \bibnamefont{Cobden}},
  \bibinfo{author}{\bibfnamefont{J.}~\bibnamefont{Lu}},
  \bibinfo{author}{\bibfnamefont{A.~G.} \bibnamefont{Rinzler}},
  \bibinfo{author}{\bibfnamefont{R.~E.} \bibnamefont{Smalley}},
  \bibinfo{author}{\bibfnamefont{L.}~\bibnamefont{Balents}}, \bibnamefont{and}
  \bibinfo{author}{\bibfnamefont{P.~L.} \bibnamefont{McEuen}},
  \bibinfo{journal}{Nature} \textbf{\bibinfo{volume}{397}},
  \bibinfo{pages}{598} (\bibinfo{year}{1999}).

\bibitem[{\citenamefont{Postma et~al.}(2001)\citenamefont{Postma, Teepen, Yao,
  Grifoni, and Dekker}}]{postma2001cns}
\bibinfo{author}{\bibfnamefont{H.}~\bibnamefont{Postma}},
  \bibinfo{author}{\bibfnamefont{T.}~\bibnamefont{Teepen}},
  \bibinfo{author}{\bibfnamefont{Z.}~\bibnamefont{Yao}},
  \bibinfo{author}{\bibfnamefont{M.}~\bibnamefont{Grifoni}}, \bibnamefont{and}
  \bibinfo{author}{\bibfnamefont{C.}~\bibnamefont{Dekker}},
  \bibinfo{journal}{Science} \textbf{\bibinfo{volume}{293}},
  \bibinfo{pages}{76} (\bibinfo{year}{2001}).

\bibitem[{\citenamefont{Liang et~al.}(2002)\citenamefont{Liang, Bockrath, and
  Park}}]{liang:prl2002}
\bibinfo{author}{\bibfnamefont{W.}~\bibnamefont{Liang}},
  \bibinfo{author}{\bibfnamefont{M.}~\bibnamefont{Bockrath}}, \bibnamefont{and}
  \bibinfo{author}{\bibfnamefont{H.}~\bibnamefont{Park}},
  \bibinfo{journal}{Phys. Rev. Lett.} \textbf{\bibinfo{volume}{88}},
  \bibinfo{pages}{126801} (\bibinfo{year}{2002}).

\bibitem[{\citenamefont{Moriyama et~al.}(2005)\citenamefont{Moriyama, Fuse,
  Suzuki, Aoyagi, and Ishibashi}}]{moriyama:prl2005}
\bibinfo{author}{\bibfnamefont{S.}~\bibnamefont{Moriyama}},
  \bibinfo{author}{\bibfnamefont{T.}~\bibnamefont{Fuse}},
  \bibinfo{author}{\bibfnamefont{M.}~\bibnamefont{Suzuki}},
  \bibinfo{author}{\bibfnamefont{Y.}~\bibnamefont{Aoyagi}}, \bibnamefont{and}
  \bibinfo{author}{\bibfnamefont{K.}~\bibnamefont{Ishibashi}},
  \bibinfo{journal}{Phys. Rev. Lett.} \textbf{\bibinfo{volume}{94}},
  \bibinfo{eid}{186806} (\bibinfo{year}{2005}).

\bibitem[{\citenamefont{Sapmaz et~al.}(2005)\citenamefont{Sapmaz,
  Jarillo-Herrero, Kong, Dekker, Kouwenhoven, and van~der
  Zant}}]{sapmaz:prb2005}
\bibinfo{author}{\bibfnamefont{S.}~\bibnamefont{Sapmaz}},
  \bibinfo{author}{\bibfnamefont{P.}~\bibnamefont{Jarillo-Herrero}},
  \bibinfo{author}{\bibfnamefont{J.}~\bibnamefont{Kong}},
  \bibinfo{author}{\bibfnamefont{C.}~\bibnamefont{Dekker}},
  \bibinfo{author}{\bibfnamefont{L.~P.} \bibnamefont{Kouwenhoven}},
  \bibnamefont{and} \bibinfo{author}{\bibfnamefont{H.~S.~J.}
  \bibnamefont{van~der Zant}}, \bibinfo{journal}{Phys. Rev. B}
  \textbf{\bibinfo{volume}{71}}, \bibinfo{eid}{153402} (\bibinfo{year}{2005}).

\bibitem[{\citenamefont{Sapmaz et~al.}(2006)\citenamefont{Sapmaz,
  Jarillo-Herrero, Kouwenhoven, and van~der Zant}}]{sapmaz2006qdc}
\bibinfo{author}{\bibfnamefont{S.}~\bibnamefont{Sapmaz}},
  \bibinfo{author}{\bibfnamefont{P.}~\bibnamefont{Jarillo-Herrero}},
  \bibinfo{author}{\bibfnamefont{L.}~\bibnamefont{Kouwenhoven}},
  \bibnamefont{and} \bibinfo{author}{\bibfnamefont{H.}~\bibnamefont{van~der
  Zant}}, \bibinfo{journal}{Semicond. Sci. Technol}
  \textbf{\bibinfo{volume}{21}}, \bibinfo{pages}{S52} (\bibinfo{year}{2006}).

\bibitem[{\citenamefont{Grabert and Devoret}(1991)}]{grabert1991set}
\bibinfo{editor}{\bibfnamefont{H.}~\bibnamefont{Grabert}} \bibnamefont{and}
  \bibinfo{editor}{\bibfnamefont{M.}~\bibnamefont{Devoret}}, eds.,
  \emph{\bibinfo{title}{{Single Electron Tunneling}}}
  (\bibinfo{publisher}{Plenum. New York}, \bibinfo{year}{1991}).

\bibitem[{\citenamefont{Alhassid}(2000)}]{alhassid:rmp2000}
\bibinfo{author}{\bibfnamefont{Y.}~\bibnamefont{Alhassid}},
  \bibinfo{journal}{Rev. Mod. Phys.} \textbf{\bibinfo{volume}{72}},
  \bibinfo{pages}{895} (\bibinfo{year}{2000}).

\bibitem[{\citenamefont{Cobden and Nyg\aa{}rd}(2002)}]{cobden:prl2002}
\bibinfo{author}{\bibfnamefont{D.~H.} \bibnamefont{Cobden}} \bibnamefont{and}
  \bibinfo{author}{\bibfnamefont{J.}~\bibnamefont{Nyg\aa{}rd}},
  \bibinfo{journal}{Phys. Rev. Lett.} \textbf{\bibinfo{volume}{89}},
  \bibinfo{pages}{046803} (\bibinfo{year}{2002}).

\bibitem[{\citenamefont{Oreg et~al.}(2000)\citenamefont{Oreg, Byczuk, and
  Halperin}}]{oreg:prl2000}
\bibinfo{author}{\bibfnamefont{Y.}~\bibnamefont{Oreg}},
  \bibinfo{author}{\bibfnamefont{K.}~\bibnamefont{Byczuk}}, \bibnamefont{and}
  \bibinfo{author}{\bibfnamefont{B.~I.} \bibnamefont{Halperin}},
  \bibinfo{journal}{Phys. Rev. Lett.} \textbf{\bibinfo{volume}{85}},
  \bibinfo{pages}{365} (\bibinfo{year}{2000}).

\bibitem[{\citenamefont{Mayrhofer and Grifoni}(2006)}]{mayrhofer:prb2006}
\bibinfo{author}{\bibfnamefont{L.}~\bibnamefont{Mayrhofer}} \bibnamefont{and}
  \bibinfo{author}{\bibfnamefont{M.}~\bibnamefont{Grifoni}},
  \bibinfo{journal}{Phys. Rev. B} \textbf{\bibinfo{volume}{74}},
  \bibinfo{eid}{121403(R)} (\bibinfo{year}{2006}).

\bibitem[{\citenamefont{Mayrhofer and Grifoni}(2007)}]{mayrhofer:epjb2007}
\bibinfo{author}{\bibfnamefont{L.}~\bibnamefont{Mayrhofer}} \bibnamefont{and}
  \bibinfo{author}{\bibfnamefont{M.}~\bibnamefont{Grifoni}},
  \bibinfo{journal}{Eur. Phys. J. B} \textbf{\bibinfo{volume}{57}},
  \bibinfo{pages}{107} (\bibinfo{year}{2007}).

\bibitem[{\citenamefont{Mayrhofer and Grifoni}()}]{mayrhofer}
\bibinfo{author}{\bibfnamefont{L.}~\bibnamefont{Mayrhofer}} \bibnamefont{and}
  \bibinfo{author}{\bibfnamefont{M.}~\bibnamefont{Grifoni}}, \bibinfo{note}{in
  preparation}.

\bibitem[{\citenamefont{Buitelaar et~al.}(2002)\citenamefont{Buitelaar,
  Bachtold, Nussbaumer, Iqbal, and Sch\"onenberger}}]{buitellar:prl2002}
\bibinfo{author}{\bibfnamefont{M.~R.} \bibnamefont{Buitelaar}},
  \bibinfo{author}{\bibfnamefont{A.}~\bibnamefont{Bachtold}},
  \bibinfo{author}{\bibfnamefont{T.}~\bibnamefont{Nussbaumer}},
  \bibinfo{author}{\bibfnamefont{M.}~\bibnamefont{Iqbal}}, \bibnamefont{and}
  \bibinfo{author}{\bibfnamefont{C.}~\bibnamefont{Sch\"onenberger}},
  \bibinfo{journal}{Phys. Rev. Lett.} \textbf{\bibinfo{volume}{88}},
  \bibinfo{pages}{156801} (\bibinfo{year}{2002}).

\bibitem[{\citenamefont{Yoon et~al.}(2002)\citenamefont{Yoon, Delaney, and
  Louie}}]{yoon:prb2002}
\bibinfo{author}{\bibfnamefont{Y.-G.} \bibnamefont{Yoon}},
  \bibinfo{author}{\bibfnamefont{P.}~\bibnamefont{Delaney}}, \bibnamefont{and}
  \bibinfo{author}{\bibfnamefont{S.~G.} \bibnamefont{Louie}},
  \bibinfo{journal}{Phys. Rev. B} \textbf{\bibinfo{volume}{66}},
  \bibinfo{pages}{073407} (\bibinfo{year}{2002}).

\bibitem[{\citenamefont{Roche et~al.}(2001)\citenamefont{Roche, Triozon, Rubio,
  and Mayou}}]{roche:prb2001}
\bibinfo{author}{\bibfnamefont{S.}~\bibnamefont{Roche}},
  \bibinfo{author}{\bibfnamefont{F.}~\bibnamefont{Triozon}},
  \bibinfo{author}{\bibfnamefont{A.}~\bibnamefont{Rubio}}, \bibnamefont{and}
  \bibinfo{author}{\bibfnamefont{D.}~\bibnamefont{Mayou}},
  \bibinfo{journal}{Phys. Rev. B} \textbf{\bibinfo{volume}{64}},
  \bibinfo{pages}{121401(R)} (\bibinfo{year}{2001}).

\bibitem[{\citenamefont{Triozon et~al.}(2004)\citenamefont{Triozon, Roche,
  Rubio, and Mayou}}]{triozon:prb2004}
\bibinfo{author}{\bibfnamefont{F.}~\bibnamefont{Triozon}},
  \bibinfo{author}{\bibfnamefont{S.}~\bibnamefont{Roche}},
  \bibinfo{author}{\bibfnamefont{A.}~\bibnamefont{Rubio}}, \bibnamefont{and}
  \bibinfo{author}{\bibfnamefont{D.}~\bibnamefont{Mayou}},
  \bibinfo{journal}{Phys. Rev. B} \textbf{\bibinfo{volume}{69}},
  \bibinfo{pages}{121410(R)} (\bibinfo{year}{2004}).

\bibitem[{\citenamefont{Wang and Grifoni}(2005)}]{wang:2005}
\bibinfo{author}{\bibfnamefont{S.}~\bibnamefont{Wang}} \bibnamefont{and}
  \bibinfo{author}{\bibfnamefont{M.}~\bibnamefont{Grifoni}},
  \bibinfo{journal}{Phys. Rev. Lett.} \textbf{\bibinfo{volume}{95}},
  \bibinfo{eid}{266802} (\bibinfo{year}{2005}).

\bibitem[{\citenamefont{Egger}(1999)}]{egger:prl1999}
\bibinfo{author}{\bibfnamefont{R.}~\bibnamefont{Egger}},
  \bibinfo{journal}{Phys. Rev. Lett.} \textbf{\bibinfo{volume}{83}},
  \bibinfo{pages}{5547} (\bibinfo{year}{1999}).

\bibitem[{\citenamefont{Saito et~al.}(1993)\citenamefont{Saito, Dresselhaus,
  and Dresselhaus}}]{saito:jap1993}
\bibinfo{author}{\bibfnamefont{R.}~\bibnamefont{Saito}},
  \bibinfo{author}{\bibfnamefont{G.}~\bibnamefont{Dresselhaus}},
  \bibnamefont{and} \bibinfo{author}{\bibfnamefont{M.~S.}
  \bibnamefont{Dresselhaus}}, \bibinfo{journal}{J. Appl. Phys.}
  \textbf{\bibinfo{volume}{73}}, \bibinfo{pages}{494} (\bibinfo{year}{1993}).

\bibitem[{\citenamefont{Uryu}(2004)}]{uryu:prb2004}
\bibinfo{author}{\bibfnamefont{S.}~\bibnamefont{Uryu}}, \bibinfo{journal}{Phys.
  Rev. B} \textbf{\bibinfo{volume}{69}}, \bibinfo{eid}{075402}
  (\bibinfo{year}{2004}).

\bibitem[{\citenamefont{Uryu and Ando}(2005)}]{uryu:245403}
\bibinfo{author}{\bibfnamefont{S.}~\bibnamefont{Uryu}} \bibnamefont{and}
  \bibinfo{author}{\bibfnamefont{T.}~\bibnamefont{Ando}},
  \bibinfo{journal}{Phys. Rev. B} \textbf{\bibinfo{volume}{72}},
  \bibinfo{eid}{245403} (\bibinfo{year}{2005}).

\bibitem[{\citenamefont{Haldane}(1981)}]{haldane1981llt}
\bibinfo{author}{\bibfnamefont{F.}~\bibnamefont{Haldane}}, \bibinfo{journal}{J.
  Phys. C} \textbf{\bibinfo{volume}{14}} (\bibinfo{year}{1981}).

\bibitem[{\citenamefont{Voit}(1994)}]{voit:rpp1994}
\bibinfo{author}{\bibfnamefont{J.}~\bibnamefont{Voit}}, \bibinfo{journal}{Rep.
  Prog. Phys.} \textbf{\bibinfo{volume}{57}}, \bibinfo{pages}{977}
  (\bibinfo{year}{1994}).

\bibitem[{\citenamefont{von Delft and Schoeller}(1998)}]{delft:adp1998}
\bibinfo{author}{\bibfnamefont{J.}~\bibnamefont{von Delft}} \bibnamefont{and}
  \bibinfo{author}{\bibfnamefont{H.}~\bibnamefont{Schoeller}},
  \bibinfo{journal}{Annalen der Physik} \textbf{\bibinfo{volume}{4}},
  \bibinfo{pages}{225} (\bibinfo{year}{1998}).

\bibitem[{\citenamefont{Giamarchi}(2004)}]{giamarchi:2004}
\bibinfo{author}{\bibfnamefont{T.}~\bibnamefont{Giamarchi}},
  \emph{\bibinfo{title}{Quantum physics in one dimension}}
  (\bibinfo{publisher}{Oxford university press}, \bibinfo{year}{2004}).

\bibitem[{\citenamefont{Matveev and Glazman}(1993)}]{matveev:prb1993}
\bibinfo{author}{\bibfnamefont{K.~A.} \bibnamefont{Matveev}} \bibnamefont{and}
  \bibinfo{author}{\bibfnamefont{L.~I.} \bibnamefont{Glazman}},
  \bibinfo{journal}{Phys. Rev. Lett.} \textbf{\bibinfo{volume}{70}},
  \bibinfo{pages}{990} (\bibinfo{year}{1993}).

\bibitem[{\citenamefont{Blum}(1996)}]{blum:1996}
\bibinfo{author}{\bibfnamefont{K.}~\bibnamefont{Blum}},
  \emph{\bibinfo{title}{Density Matrix Theory and Applications}}
  (\bibinfo{publisher}{Plenum Press, New York}, \bibinfo{year}{1996}).

\bibitem[{\citenamefont{Bloch}(1957)}]{bloch:pr1957}
\bibinfo{author}{\bibfnamefont{F.}~\bibnamefont{Bloch}},
  \bibinfo{journal}{Phys. Rev.} \textbf{\bibinfo{volume}{105}},
  \bibinfo{pages}{1206} (\bibinfo{year}{1957}).

\bibitem[{\citenamefont{Redfield}(1957)}]{redfield1957trp}
\bibinfo{author}{\bibfnamefont{A.}~\bibnamefont{Redfield}},
  \bibinfo{journal}{IBM J. Res. Dev} \textbf{\bibinfo{volume}{1}},
  \bibinfo{pages}{19} (\bibinfo{year}{1957}).

\bibitem[{\citenamefont{Glazman and Matveev}(1988)}]{glazman:jetpl1988}
\bibinfo{author}{\bibfnamefont{L.~I.} \bibnamefont{Glazman}} \bibnamefont{and}
  \bibinfo{author}{\bibfnamefont{K.~A.} \bibnamefont{Matveev}},
  \bibinfo{journal}{JETP Lett} \textbf{\bibinfo{volume}{48}},
  \bibinfo{pages}{445} (\bibinfo{year}{1988}).

\bibitem[{\citenamefont{Beenakker}(1991)}]{beenakker:prb1991}
\bibinfo{author}{\bibfnamefont{C.~W.~J.} \bibnamefont{Beenakker}},
  \bibinfo{journal}{Phys. Rev. B} \textbf{\bibinfo{volume}{44}},
  \bibinfo{pages}{1646} (\bibinfo{year}{1991}).

\bibitem[{\citenamefont{Gradshteyn and Ryzhik}(2000)}]{gradshteyn:2000}
\bibinfo{author}{\bibfnamefont{I.~S.} \bibnamefont{Gradshteyn}}
  \bibnamefont{and} \bibinfo{author}{\bibfnamefont{I.}~\bibnamefont{Ryzhik}},
  \emph{\bibinfo{title}{Table of Integral, Series and Products}}
  (\bibinfo{publisher}{Academic Press, San Diego}, \bibinfo{year}{2000}).

\end{thebibliography}
\end{document}